\documentclass[12pt]{iopart}
\usepackage{iopams}
\usepackage{graphicx}
\usepackage{subfig}
 \pdfoutput=1

\usepackage[version=0.96]{pgf}

\newcommand{\bsigm}{{\boldsymbol \sigma}}

\begin{document}

\title[Belief-Propagation and replicas for learning]{Belief-Propagation and replicas for inference and learning in a kinetic Ising model with hidden spins}

\author{C Battistin$^1$, J Hertz$^{2,3}$, J Tyrcha$^4$, Y Roudi$^{1,2}$}

\address{$^1$ Kavli Institute for Systems Neuroscience and Centre for Neural Computation , Olav Kyrres gate 9, 7030 Trondheim, NO}
\address{$^2$ NORDITA, KTH Royal Institute of Technology and Stockholm University, 10691 Stockholm, Sweden}
\address{$^3$ Niels Bohr Institute, Blegdamsvej 17, 2100 Copenhagen Ø, Denmark}
\address{$^4$ Mathematical Statistics, Stockholm University, S106 91 Stockholm, Sweden}
\ead{claudia.battistin@ntnu.no}

\vspace{10pt}
\begin{indented}
\item[]December 2014
\end{indented}

\begin{abstract}
We propose a new algorithm for inferring the state of hidden spins and reconstructing the connections in a synchronous kinetic Ising model,  given the observed history. Focusing on the case in which the hidden spins are conditionally independent of each other given the state of observable spins, we show that calculating the likelihood of the data can be simplified by introducing a set of replicated auxiliary spins. Belief Propagation (BP) and Susceptibility Propagation (SusP) can then be used to infer the states of hidden variables and learn the couplings. We study the convergence and performance of this algorithm for networks with both Gaussian-distributed and binary bonds. We also study how the algorithm behaves as the fraction of hidden nodes and the amount of data are changed, showing that it outperforms the TAP equations for reconstructing the connections. 
\end{abstract}

%
%
%
%
%

\section{Introduction}

Reconstructing interactions in a complex network and building statistical models for their stochastic dynamics are important and active areas of research in statistical physics  and machine learning. In recent years, questions related to this topic have received extra attention given the applications in analyzing high-throughput biological (e.g. gene expression or neural) as well as financial data. The statistical physics community has contributed significantly to this area in recent years offering tools to implement approximate and efficient inference algorithms as well as theoretical insight into various aspects of the problem \cite{lezon2006using,braunstein2008inference,cocco2009neuronal,bury2013market,roudi2009ising}. 

Although the amount of data that one can record from, e.g. biological networks, is rapidly increasing, even in the best cases we still do not have access to recordings from the whole network. For example, with the most advanced recording technologies today, we can still only measure the activity of a fraction of neurons in a local cortical circuit, or the expression level of a fraction of genes or proteins in a gene regulatory or protein-protein interaction network. This raises the challenge of how to infer network connectivity and how to build statistical models of this data in the presence of hidden nodes. 
As a generic and practically useful platform for studying inference from kinetic high-throughput data, we focus our attention in this paper on a partially observed spin system: the kinetic Ising model with parallel update dynamics in the presence of hidden units. 

With asynchronous updating and symmetric connectivity, the kinetic Ising model approaches an equilibrium Gibbs distribution. Statistical physicists have contributed significantly to building approximate learning methods for the equilibrium Ising model. Among them the most studied are Na\"{i}ve Mean Field \cite{tanaka1998mean,kholodenko1990onsager}, Thouless-Anderson-Palmer approximation (TAP, i.e., first and second order Plefka expansion) \cite{thouless1977solution} and a message passing algorithm called Belief Propagation \cite{bethe1935statistical,mezard2001bethe,mezard2003cavity}. Nevertheless the assumption of symmetric connectivity in biological networks is not realistic, and making it can lead to incorrect identification of interactions. Consequently, the analysis of kinetic Ising models without symmetry in the couplings has recently attracted attention \cite{Roudi:2011fk,mezard2011exact,aurell2011message,aurell2012dynamic,roudi2011dynamical,Saad2014General} and has led to exact as well as approximate methods for network reconstruction and prediction of the network dynamics. In particular, mean-field and TAP-like approximations have also been adapted to cases with hidden nodes, i.e., when the spin history is observed only for some of the spins and one is asked to reconstruct the connectivity given only these partial data \cite{dunn2013learning,hertz2014Network}. The performance of an optimal decoder for inferring the state of hidden spins has also been recently studied in \cite{romano2014Infer}.

The maximum-likelihood reconstruction of networks with hidden nodes can be worked out by \textit{Expectation Maximization} (EM) \cite{sundberg1974maximum,dempster1977maximum}, a two-step recursive algorithm that alternates between computing hidden moments at fixed connectivity and updating the connectivity given the moments. In the case of kinetic Ising model with hidden nodes, as shown in \cite{dunn2013learning,hertz2014Network}, this EM algorithm can be done approximately using mean-field and TAP approximations: approximations that can be thought of as small coupling expansions. Another class of approximations, more appropriate for sparse and strong connectivity, similar to some biological networks, is cavity/Belief-Propagation (BP) approximations. For the kinetic Ising model without hidden nodes, these approximations have been studied in \cite{neri2009cavity,aurell2011three}. In this paper, we focus our attention to the inference of hidden states and learning network connectivity with hidden nodes, using BP and its extension to calculating susceptibilities, Susceptibility Propagation (SusP) in a replica-based approach. We derive an algorithm based on these approximations for the case where there are no hidden-to-hidden connections and numerically evaluate its performance under different conditions. 

The paper is organized as follows. In the first section we define the dynamical model and introduce a replicated system of auxiliary hidden nodes that simplifies calculating the likelihood. We then derive the update rules for the couplings in a gradient-ascent fashion on likelihood, emphasizing the required expectation values to be taken over the distribution of hidden node values. These are expressed in terms of cavity messages, which in turn obey a closed set of equations parametrized by the couplings. In the second section, we test our EM\- protocol on two architectures and compare its performance with that of other mean-field methods. 

\section{Formulation of the model}\label{sec:Form}

We consider a network of $N=N_\mathrm{v}+N_\mathrm{h}$ binary $\pm1$ spins . We observe only some of them, distinguishing between $N_\mathrm{v}$ visible units, whose states at time $t$ we denote as $\bi{s}(t)=\{s_i(t)\}$, and  $N_\mathrm{h}$ hidden units, with states $\bsigma(t)=\{\sigma_a(t)\}$. The dynamics is Markovian and the spins are updated synchronously at each time step, according to the transition probability:

\begin{equation}
\fl{\rm P}\left[ \bi{s}(t+1),\bsigm(t+1)\vert \bi{s}(t),\bsigm(t)\right]= \frac{\exp\left[ \sum_i s_i(t+1)h_i(t)\right] }{\prod_i 2 \cosh\left[ h_i(t)\right] }\times \frac{\exp\left[ \sum_a \sigma_a(t+1)b_a(t)\right] }{\prod_a 2 \cosh\left[ b_a(t)\right] },
\label{eq:dynamics}
\end{equation}
where we choose the fields acting on visible and hidden units respectively as follows: 

\begin{eqnarray}
h_i(t)&=&\sum_j J_{ij}s_j(t)+ \sum_b K_{ib} \sigma_b(t) \label{eq:JtoVisible}\\
b_a(t)&=&\sum_j L_{aj}s_j(t) \label{eq:JtoHidden}.
\end{eqnarray}
Notice that the units are conditionally independent given the state of the network at the previous time step and that we restrict ourselves to a system without hidden-to-hidden interactions. This choice is important for the following derivation. 
We do not make any further assumption on the connectivity; in particular, the couplings do not need to be either symmetric or fully asymmetric.  We observe, however, that because there are no hidden-to-hidden connections the likelihood factorizes in time ${\rm P}\left[ \bi{s}\right]=\prod_t {\rm P}_t\left[ \bi{s}\right]$, where each factor involves a trace over single-time hidden states $\bsigma(t)$:
\begin{equation}
\fl {\rm P}_t\left[ \bi{s}\right]=\sum_\bsigm \frac{\exp \left[ \sum_i s_i^+ \left( \sum_j J_{ij}s_j + \sum_b K_{ib}\sigma_b\right) +\sum_{a,j} \sigma_a L_{aj}s_j^- \right] }{\prod_i 2\cosh\left[ \sum_j J_{ij}s_j + \sum_b K_{ib}\sigma_b\right] \prod_a 2\cosh \left[  \sum_j L_{aj}s_j^-\right]  }\qquad .
\label{eq:factLike}
\end{equation}
Here we have dropped the argument $t$ on the $\sigma$s and written $s_j^{\pm}$ for $s_j (t \pm 1)$. 

Suppose that we record the states of the visible units in the time-interval $[1,T]$. The likelihood of this history under the stochastic process in (\ref{eq:dynamics}) is obtained by tracing out the hidden states from the joint distribution of the histories of hidden and visible states. Given the $J$s, $K$s and $L$s this operation is exponentially complex in $N_{\rm h}$ and can therefore only be performed exactly for very small systems (in practice $N_{\rm h}$ of order 10).  For larger networks, approximate methods, such as those in \cite{dunn2013learning} and \cite{hertz2014Network}, have to be introduced.  In this paper, we introduce an approximation scheme that uses the BP and SusP algorithms.  To formulate it, we will need to introduce replicas.

\subsection{Introducing the replicas}

On the right-hand side of (\ref{eq:factLike}) we are dealing with a system of hidden units interacting with future, past and present observations. From now on we are going to treat our observations as sources of an external field, rather than states of reacting units. One can then notice that the only term that makes the distribution we are integrating over different from the usual equilibrium Ising distribution for the hidden units is the first hyperbolic cosine in the denominator.

Consider now the following equality:

\begin{equation}
\left(  2\cosh\left[ f_i\right]\right)^n= \sum_{\btau_i} \exp \left[ \sum_{\alpha=1}^n  \tau_i^\alpha  f_i \right] 
\label{eq:replica} 
\end{equation}
with $f_i=\sum_j J_{ij}s_j + \sum_b K_{ib}\sigma_b$, that introduces a trace over $n$ replicas of $N_\mathrm{v}$ auxiliary spins $\tau^{\alpha}_i=\pm 1$. Defining 

\begin{equation}
\fl {\rm P}^{(n)}_t\left[ \bi{s} \right]\equiv \sum_{\bsigm,\btau} \frac{\exp \left[ \sum_i \left( s_i^+ + \sum_{\alpha=1}^n \tau_i^\alpha\right) \left( \sum_j J_{ij}s_j + \sum_b K_{ib}\sigma_b\right) +\sum_{a,j} \sigma_a L_{aj}s_j^- \right] }{\prod_a 2\cosh \left[  \sum_j L_{aj}s_j^-\right]  },
\label{eq:factLikeRep}
\end{equation}
using Eq.\ (\ref{eq:replica}) and then taking the limit of the number of replicas to $-1$, for the likelihood of the data we have

\begin{equation}
\fl {\rm P}_t\left[ \bi{s} \right]= \lim_{n\to {-1}}{\rm P}^{(n)}_t\left[ \bi{s} \right] \label{eq:limitlike}
\end{equation}

In the next section, using the derivatives of ${\rm P}^{(n)}$ we will derive learning rules for the couplings. Assuming that the order of these derivatives and the limit of $n \to -1$ can be exchanged, we can then obtain learning rules for the system with hidden nodes by taking the $n\to -1$ limit. 

\begin{figure}
\centering
               \includegraphics[width=150mm,height=120mm]{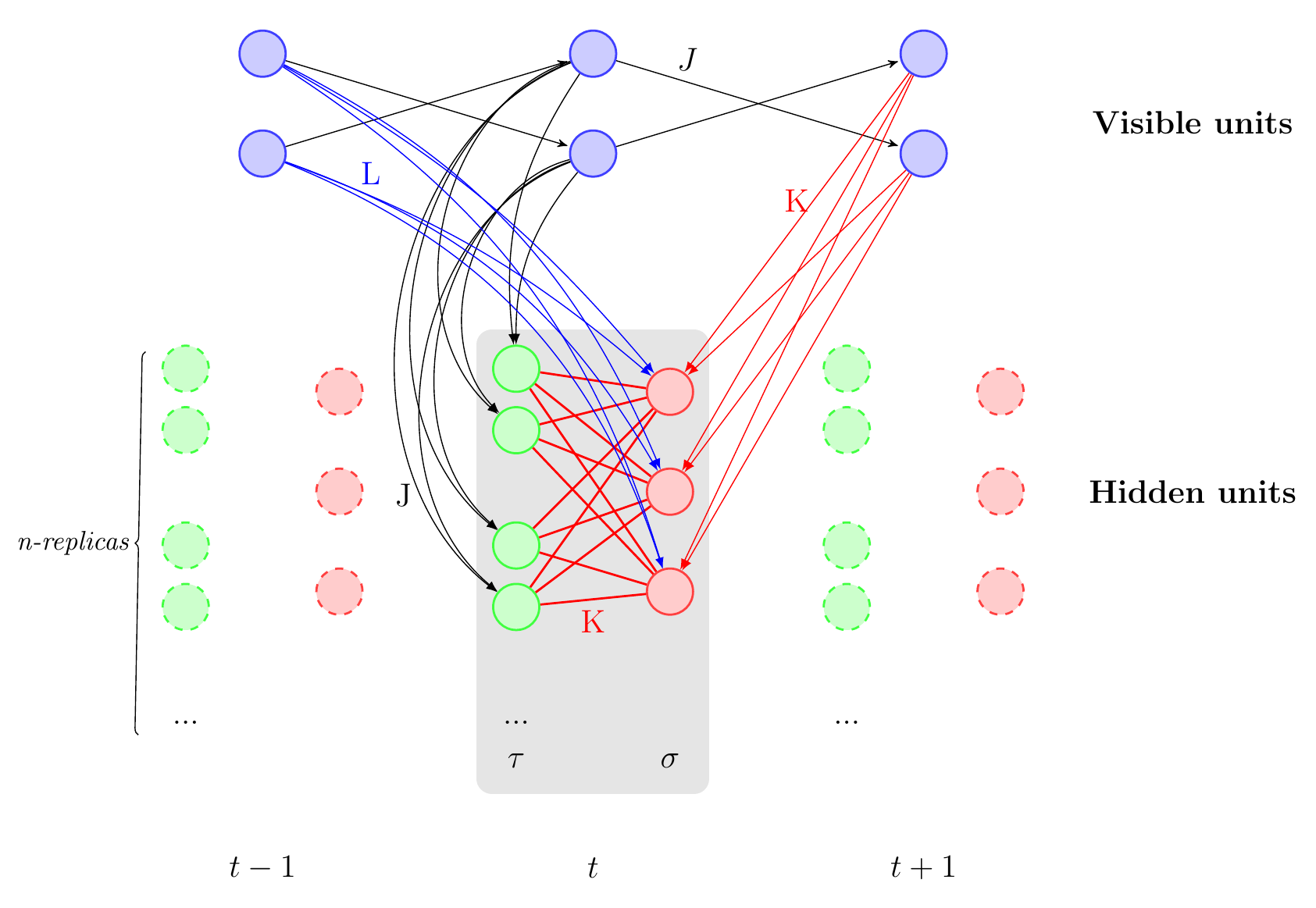}
\caption{\label{fig:replica}Time slicing of the dynamics. Equilibrium bipartite system of hidden nodes $\sigma$s and $n$ replicas of auxiliary hidden nodes at time t, interacting via $K$s. Observed nodes at $t-1$, $t$ and $t+1$ provide a different external input at each time step.}
\end{figure}

Notice that every auxiliary hidden unit $\tau_i^{\alpha}$ feels the field $\sum_j J_{ij}s_{j}$ generated by the data and couples to the $\sigma_b$ via the $K_{ib}$. These couplings are now the only interaction left in the problem: everything else can just be thought of as external fields, as far as the $\sigma$s and $\tau$s are concerned. We can therefore now use standard methods for Ising spin systems on this rather conventional (except for having $-1$ species of the $\tau$ spins) problem. Figure \ref{fig:replica} illustrates the new system.

\subsection{Learning rules}

The EM algorithm provides a simple and robust tool for parameter estimation in models with incomplete data \cite{wu1983convergence}. In the maximization step, the model parameters are updated in order to maximize the likelihood given the statistics of the hidden variables. Here we implement this step using the gradient based learning rule $\Delta J_{ij} \propto \partial \log {\rm P}(\bi{s})/\partial J_{ij}$. Using  (\ref{eq:factLikeRep}) and (\ref{eq:limitlike}) one gets
\begin{equation}
\Delta J_{ij} \propto \lim_{n\to -1} \left( s_i^+ +\sum_{\alpha=1}^n \left\langle \tau_i^\alpha\right\rangle \right)  s_j,
\end{equation}
where the average denoted by $\langle \cdots \rangle$ is taken over the conditional distribution ${\rm P}^{(n)}(\bsigma,\btau\vert \bi{s} )={\rm P}^{(n)}(\bsigma,\btau, \bi{s} )/ {\rm P}^{(n)}(\bi{s} )$. (This is, of course, the contribution to $\Delta J_{ij}$ from a single time step; the full $\Delta J_{ij}$ is obtained by summing over steps.  In what follows this summation will be implicit.)

In the replica-symmetric Ansatz the replicated systems are indistinguishable and consequently the magnetizations $\left\langle \tau_i^\alpha\right\rangle$ are all equal for $\alpha=1\dots n$. Denoting their common value by  $m_i$, that is,

\begin{equation}
m_i\equiv \langle \tau^{1}_i \rangle=\langle \tau^{2}_i \rangle=\cdots =\langle \tau^{n}_i \rangle
\end{equation}
and taking the limit $n\to-1$, we have
\numparts
\begin{eqnarray}
\Delta J_{ij} &\propto & s_i^+ s_j -m_is_j \label{eq:LearningJ}\\
\Delta K_{ib} &\propto & s_i^+ \mu_b -\left\langle \tau_i \sigma_b\right\rangle \label{eq:LearningK} \\
\Delta L_{aj} &\propto & \mu_a s_j^- - \tanh\left[ \sum_b L_{bj}s_j^-\right] s_j^-,
\label{eq:LearningL}
\end{eqnarray}
\endnumparts
where $\mu_a\equiv \langle \sigma_a \rangle$ and for deriving $\Delta K_{ib}$ and $\Delta L_{aj}$ we followed the same procedure as exemplified for the $\Delta J_{ij}$. From (\ref{eq:LearningJ})-(\ref{eq:LearningL}), we see that for performing the expectation step, we need to evaluate the  mean values of $\tau$ and $\sigma$, i.e. $m_i$ and $\mu_a$, and the pair averages $\left\langle \tau_i \sigma_b\right\rangle $.  

These learning rules are exact and could have been derived without replicas:  With the identification $\tau_i = \tanh f_i$, they are just those of ref.\ \cite{hertz2014Network}, Eqs.\ (23-26).   What requires approximation (except for very small systems) is evaluating the averages.  In the next subsection, we will show how they can be calculated approximately using the BP and SusP algorithms, within the replica framework.  

\subsection{Calculating $m_i$, $\mu_a$ and $\left\langle \tau_i \sigma_b\right\rangle$}
Consider first the marginals $m_{i}$ and $\mu_a$. These can be obtained from the cavity magnetizations $m_{i\alpha}^{a}$ and $\mu_a^{i\alpha}$, which are the magnetizations calculated as if the connectivity is tree-like, with the connection from the unit labeled by the superscript removed. Standard cavity arguments \cite{mezard2001bethe} lead to the following closed set of equations for the cavity magnetizations

\numparts
\begin{eqnarray}
\mu_a^{i}&=&\tanh\left[ g_a-\sum_{j}\tanh^{-1}\left[ t_{ja}m_{j}^a\right] -\tanh^{-1}\left[ t_{ia}m_{i}^a\right] \right] \label{eq:cavityREP1}\\
m_{ i}^a&=&\tanh\left[ \sum_j J_{ij}s_j+\sum_{b\neq a} \tanh^{-1}\left[ t_{ib}\mu_b^{i}\right] \right]\label{eq:cavityREP2} 
\end{eqnarray}
\endnumparts
where we have defined $g_a=\sum_i K_{ia}s_i^+ +\sum_j L_{aj}s_j^-$ and $t_{ja}=\tanh\left[ K_{ja}\right]$. In \eref{eq:cavityREP1}-\eref{eq:cavityREP2} we got rid of the replica index $\alpha$, since by replica symmetry $\mu_a^{i\alpha}$ is independent of $\alpha$, hence the limit $n\to-1$ can be easily taken. In order to retrieve the magnetization $m_i$ one has to reintroduce the removed bond, through the relation $\tanh^{-1}[m_i]=\tanh^{-1}[m^a_i]+\tanh^{-1}[t_{ia}\mu_a^i]$, and similarly for the $\mu_a$. 

How to estimate the $\left\langle \tau_i \sigma_b\right\rangle $?
Inspired by the Susceptibility Propagation algorithm \cite{mezard2009constraint,aurell2010dynamics} we exploit the fluctuation-response theorem that matches the correlation between two nodes with the response of the magnetization at one site to a perturbation at the other site. In our problem it translates into the relation   
\begin{eqnarray}
\left\langle \tau_i \sigma_a\right\rangle =\chi_{ib}+m_i\mu_a\\
\chi_{ib}= \frac{\partial m_i}{\partial b_b^-}
\end{eqnarray}
where $\chi_{ib}$ is the susceptibility. Defining 
\begin{equation}
b_a^-\equiv\sum_j L_{aj}s_j^-,
\end{equation}
we derive a closed set of equations for the derivatives $v_{ab}^i$  and $g_{jb}^a$ of the cavity fields: 

\numparts
\begin{eqnarray}
v_{ab}^i&\equiv&  \frac{\partial \mu_a^i}{\partial b_b^-}=\left( 1-(\mu_a^{i})^2\right)\left\lbrace \delta_{ab}-\sum_{j} \frac{t_{ja}}{1-t_{ja}^2}g_{jb}^a\right\rbrace \label{eq:cavityGRAD1}\\
g_{jb}^a&\equiv& \frac{\partial m_j^a}{\partial b_b^-}=\left( 1-(m_j^{a})^2\right)\sum_{c\neq a}\frac{t_{jc}}{1-t_{jc}^2 (\mu_c^{j})^2}v_{cb}^j  
\label{eq:cavityGRAD2}
\end{eqnarray}
\endnumparts
From the stationary points of \eref{eq:cavityGRAD1}-\eref{eq:cavityGRAD2} one can recover the susceptibility using:

\begin{equation}
\chi_{ib}=\left( 1-m_i^2\right)\sum_a \frac{t_{ia}}{1-t_{ia}^2(\mu_a^i)^2}v_{ab}^i
\end{equation}

\section{Numerical Results}

We tested the reconstruction power of our algorithm on two bond distributions: Gaussian and binary. In both cases we investigated the impact of the sparsness on the performance of the algorithm, by varying the sparsity parameter $c={\rm Pr}\left[ \textrm{node $i$ is connected to node $j$}\right] $.

\subsection{Gaussian bounds}

Once we have decided which nodes are connected to which (which bonds to be non-zero, an event that happens with probability $c$), we choose the non-zero couplings to be independently drawn from a normal distribution with zero mean and standard deviation:

\numparts
\begin{eqnarray}
\sigma_{\rm J}  & = & g/\sqrt{2N_{\rm v} c} \label{eq:stdJs}\\
\sigma_{\rm L} & = & g/\sqrt{N_{\rm v} c} \label{eq:stdLs}\\
\sigma_{\rm K} & = & g/\sqrt{2N_{\rm h} c} \label{eq:stdKs}
\end{eqnarray} 
\endnumparts
where $g$ is the couplings strength parameter.
This choice ensures the field exerted on hidden and visible units to be of the same size, despite the absence of hidden to hidden interactions.

\subsubsection{Inferred statistics}

\begin{figure}[h]
        \centering
               \includegraphics[width=50mm,height=50mm]{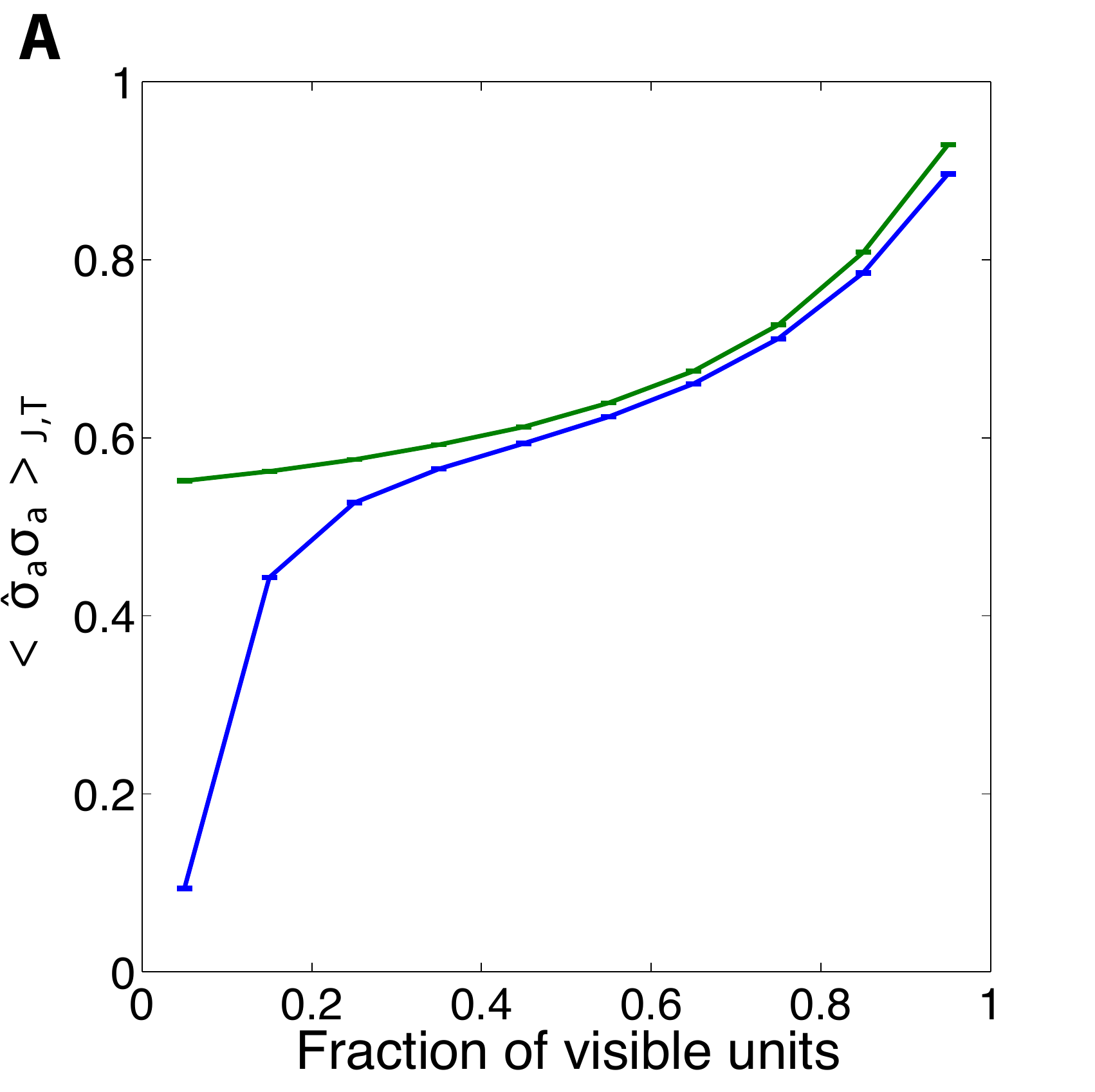}
                \label{fig:InfBP}
                \includegraphics[width=50mm,height=50mm]{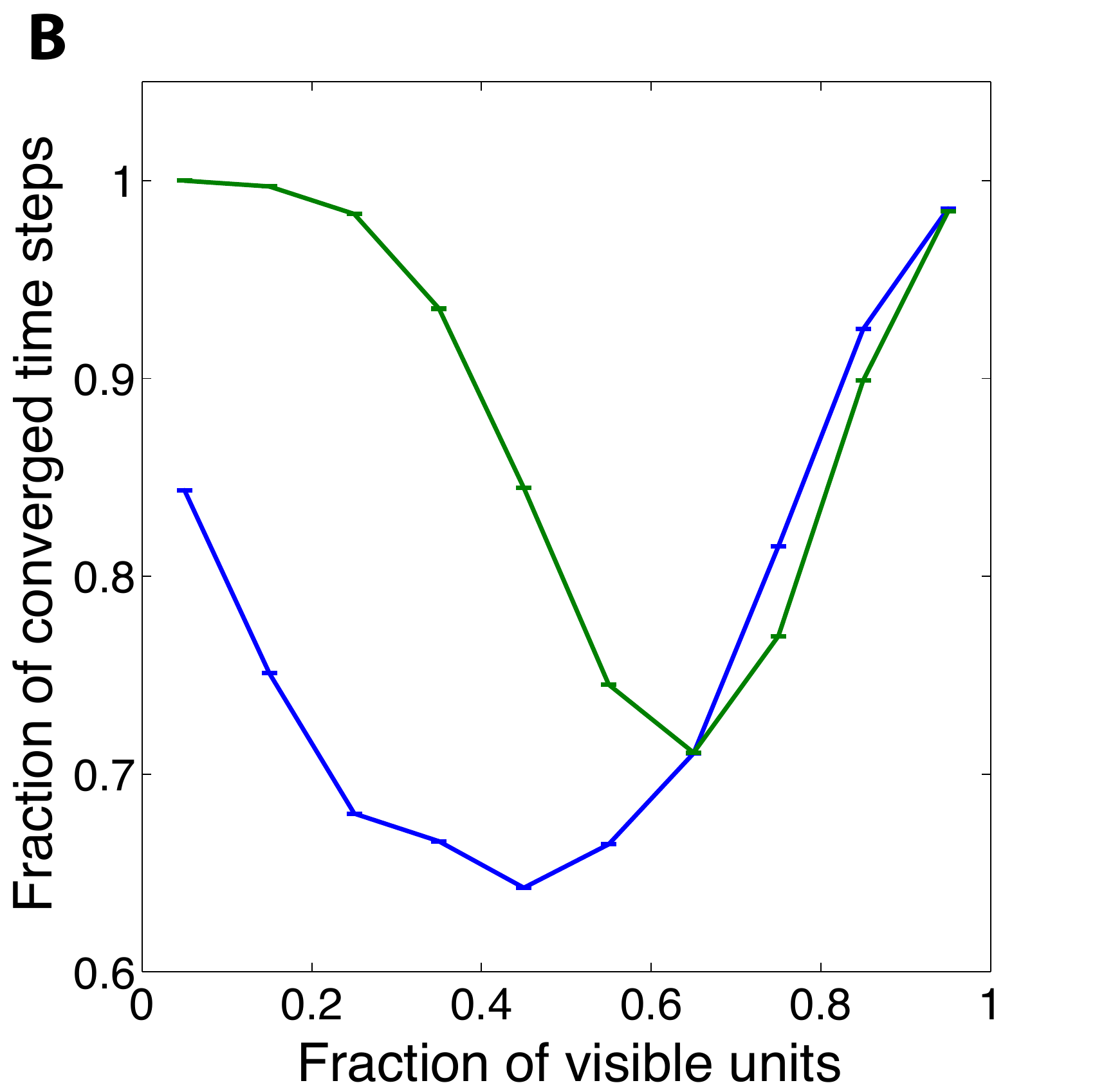}
                \label{fig:CONV(BP)}\\
	\includegraphics[width=50mm,height=50mm]{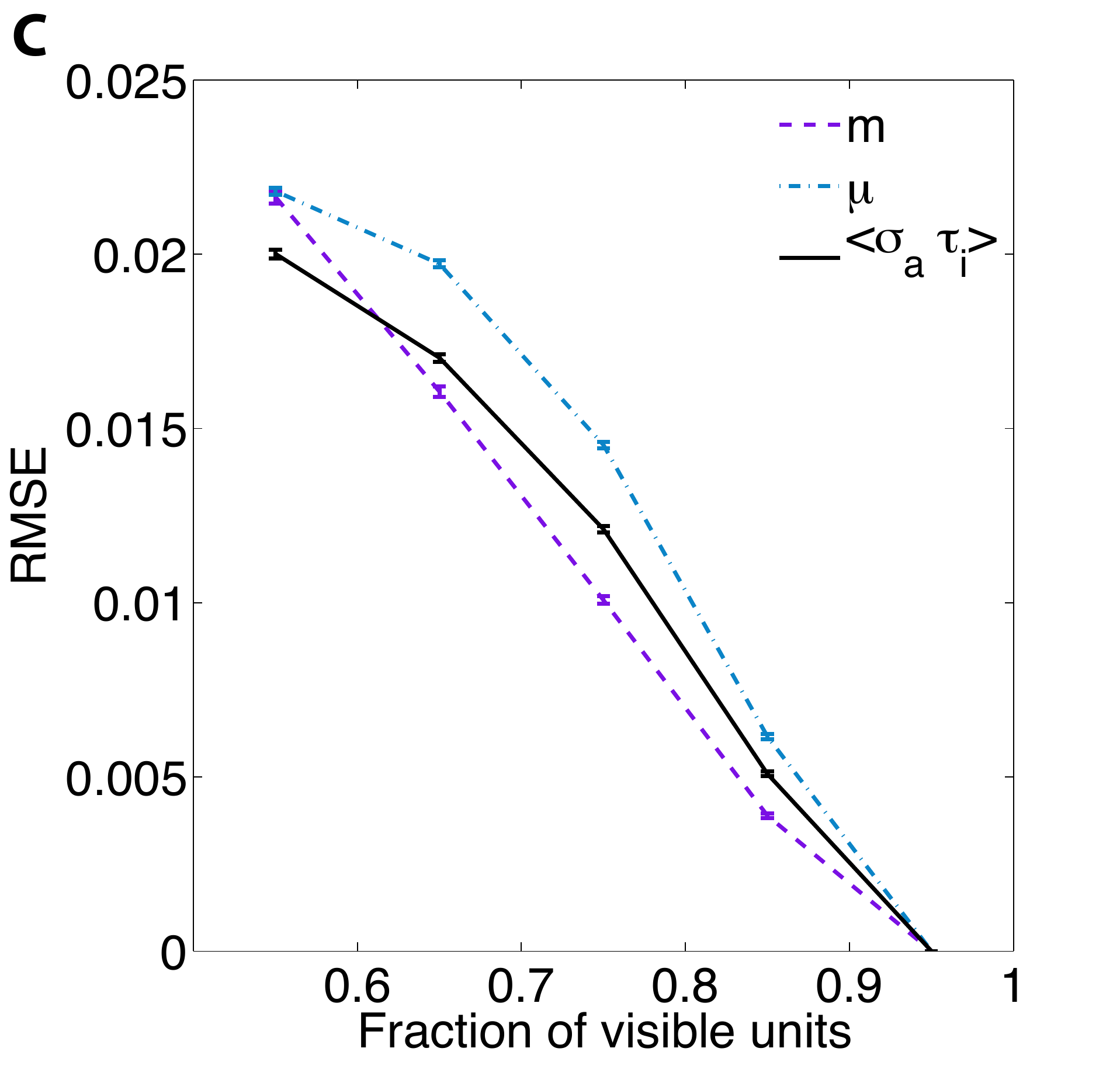}
        \label{fig:RMSE(BP-Replica)T=10}
        \caption{BP performances on inferring the statistics of hidden units vs fraction of visible units. (A) Correlation between hidden states and sign of the inferred magnetizations $\hat{\sigma}_a(t)$.  (B) fraction of converged instances. $N=100$, $T=100$, averages over $10^5$ realizations of the couplings, sparse graphs with c=0.1 (blue) and fully connected ones (green). (C) RMSE between BP and exact gradient ascent (equations (\ref{eq:ExactM})-(\ref{eq:ExactCorr})) magnetizations and correlation functions. $N=20$, $T=10$, averages over $10^6$ realizations of the couplings, $c=0.1$. }\label{fig:CONVBP}
\end{figure}

In section \ref{sec:Form} we set up our learning protocol in two steps: expectation and maximization. In the expectation (E) step we estimate means and correlations of the hidden units given observations and the current estimate of the couplings.  Here we check the performance of the E step, i.e., equations (\ref{eq:cavityREP1})-(\ref{eq:cavityREP2}) and (\ref{eq:cavityGRAD1})-(\ref{eq:cavityGRAD2}), given the true connectivity. First we study the convergence of the BP equations and the correlation between the hidden states $\sigma_a(t)$ and the estimator $\hat{\sigma}_a(t)\equiv{\rm sign}\left( \mu_a(t)\right) $.

Figure \ref{fig:CONVBP}A shows that the correlation $\left\langle\sigma_a(t)\hat{\sigma}_a(t) \right\rangle $ increases monotonically with the fraction of visible units on both on fully-connected and sparse graphs. The change of concavity at $N_{\rm v}=N_{\rm h}$ has to be ascribed to the fact that more visible units correspond to more information on the state of hidden units which make inference simpler, but also to more loops, which makes Belief Propagation algorithm less accurate. Figure \ref{fig:CONVBP}B seems to indicate that our algorithm has better performance on fully connected graphs than on sparse ones when $N_{\rm v}<N_{\rm h}$. We would  like, however, to point out that for such a small and highly dilute random network, hidden units have few connections with visible units and this can contribute to the low performance at  $N_{\rm v}<N_{\rm h}$ for sparse networks. 

For small systems, we were able to compare our BP-based results with ``exact'' ones (i.e., the best possible ones, given the finite data set used to base the inference on).  Recall that the our gradient ascent learning rules (\ref{eq:LearningJ})\--(\ref{eq:LearningL}) are exact when the averages $m_i$, $\mu_b$ and $\langle \tau_i \sigma_b \rangle$ are evaluated using the true distribution ${\rm P}_t\left[ \bsigm \vert \bi{s}\right]$:
\numparts
\begin{eqnarray}
m_i&=&{\rm Tr}_{\bsigm}\left\lbrace  \tanh\left[ \sum_j J_{ij}s_j + \sum_b K_{ib}\sigma_b\right]{\rm P}_t\left[ \bsigm \vert \bi{s}\right] \right\rbrace  \label{eq:ExactM}\\
\mu_b &=&{\rm Tr}_{\bsigm} \left\lbrace \sigma_b{\rm P}_t\left[ \bsigm \vert \bi{s}\right]\right\rbrace   \\
\left\langle \tau_i \sigma_b\right\rangle &=&{\rm Tr}_{\bsigm} \left\lbrace \tanh\left[ \sum_j J_{ij}s_j + \sum_b K_{ib}\sigma_b\right]\sigma_b{\rm P}_t\left[ \bsigm \vert \bi{s}\right] \right\rbrace \label{eq:ExactCorr}
\end{eqnarray} 
\endnumparts  
These averages can be computed by direct summation over states for small systems.  Comparing the
BP-based estimates of the parameters with those obtained in this way enabled us to see how much of the error we found (relative to ground truth) was due to BP (Figure\ \ref{fig:CONVBP}C) and how much was due to the finiteness of the data set.

If one removes the cases for which SusP diverged, the RMSE between the correlations, $\left\langle \sigma_a\tau_i\right\rangle $, inferred using BP and those inferred using exact enumeration drops as the fraction of visible units increase. In the divergent case, typically,  correlations inferred using BP grow exponentially while iterating (\ref{eq:cavityGRAD1})-(\ref{eq:cavityGRAD2}).  The lack of convergence of SusP has been already pointed out in the literature \cite{aurell2010dynamics}, and \cite{ricci2012bethe} suggests an alternative formula to compute the two point correlation function that provides finite estimates, without solving the issue, however. In the next section we study the convergence of the learning algorithm and propose a simple but effective method to estimate the correlations between hidden units.         

\subsubsection{Learning}

\begin{figure}[h]
        \centering

        \includegraphics[width=40mm,height=40mm]{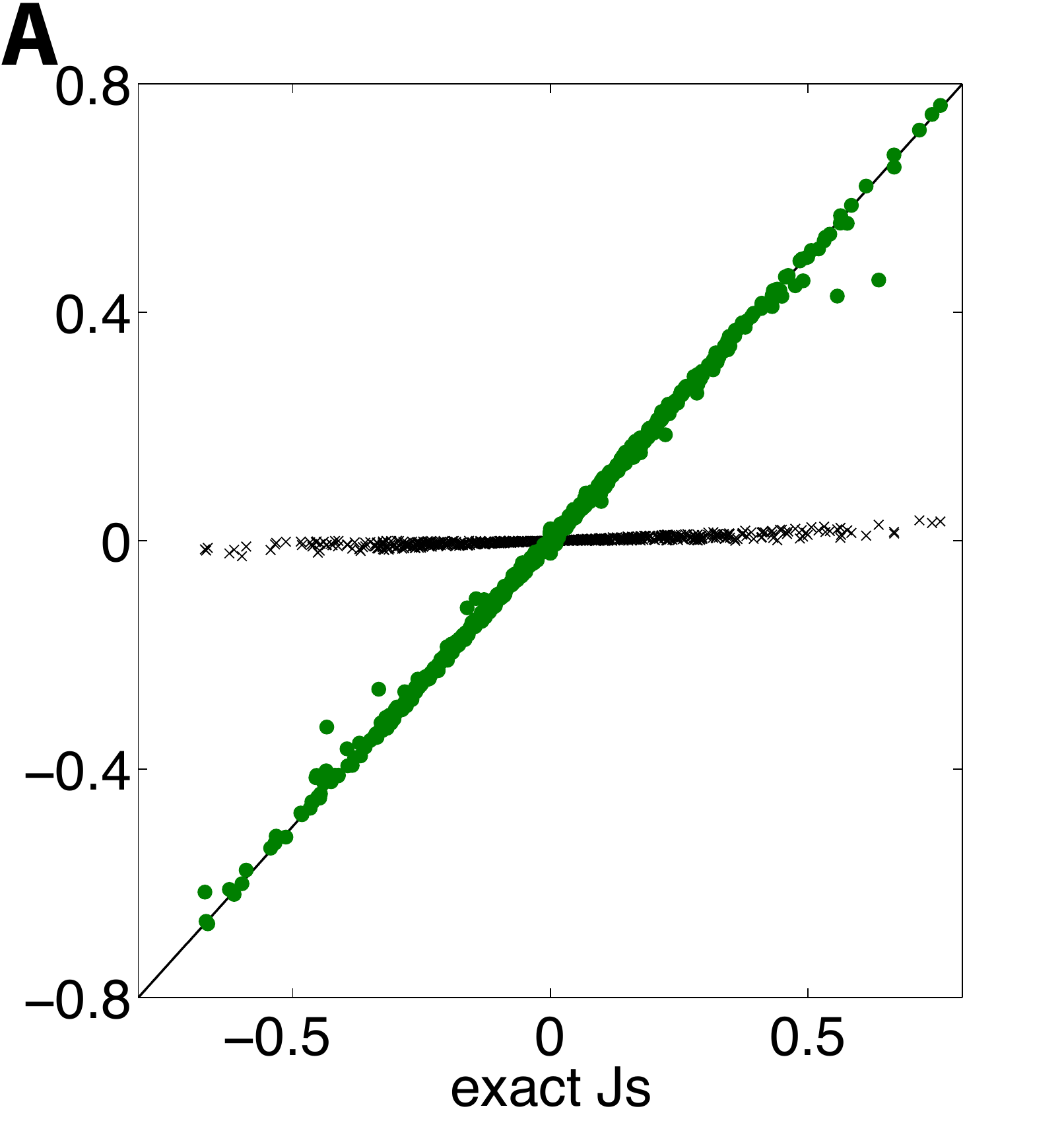}
                \label{fig:Js}
                \includegraphics[width=40mm,height=40mm]{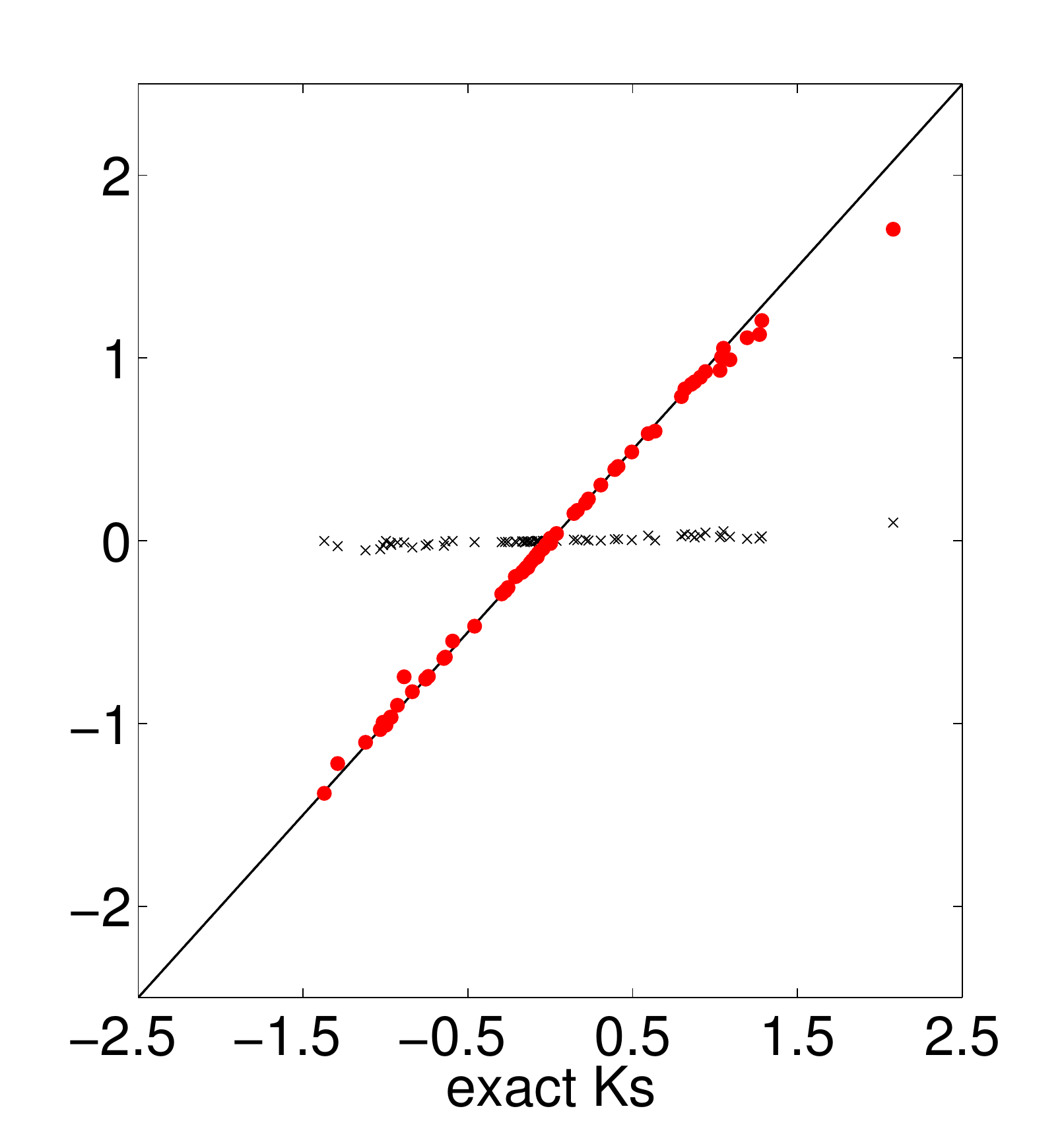}
                \label{fig:Ks}
                \includegraphics[width=40mm,height=40mm]{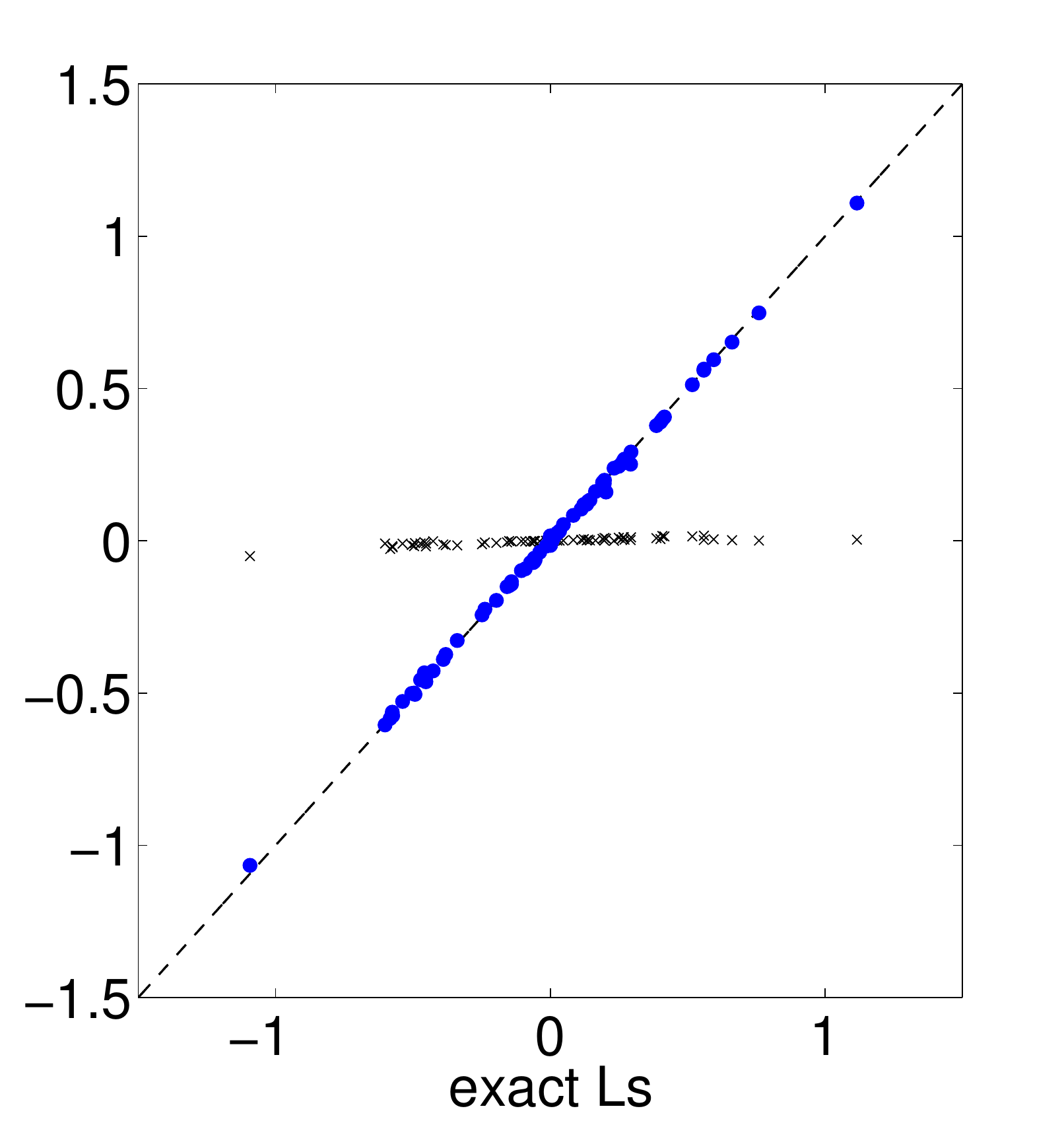}
                \label{fig:Ls}
                \includegraphics[width=60mm,height=60mm]{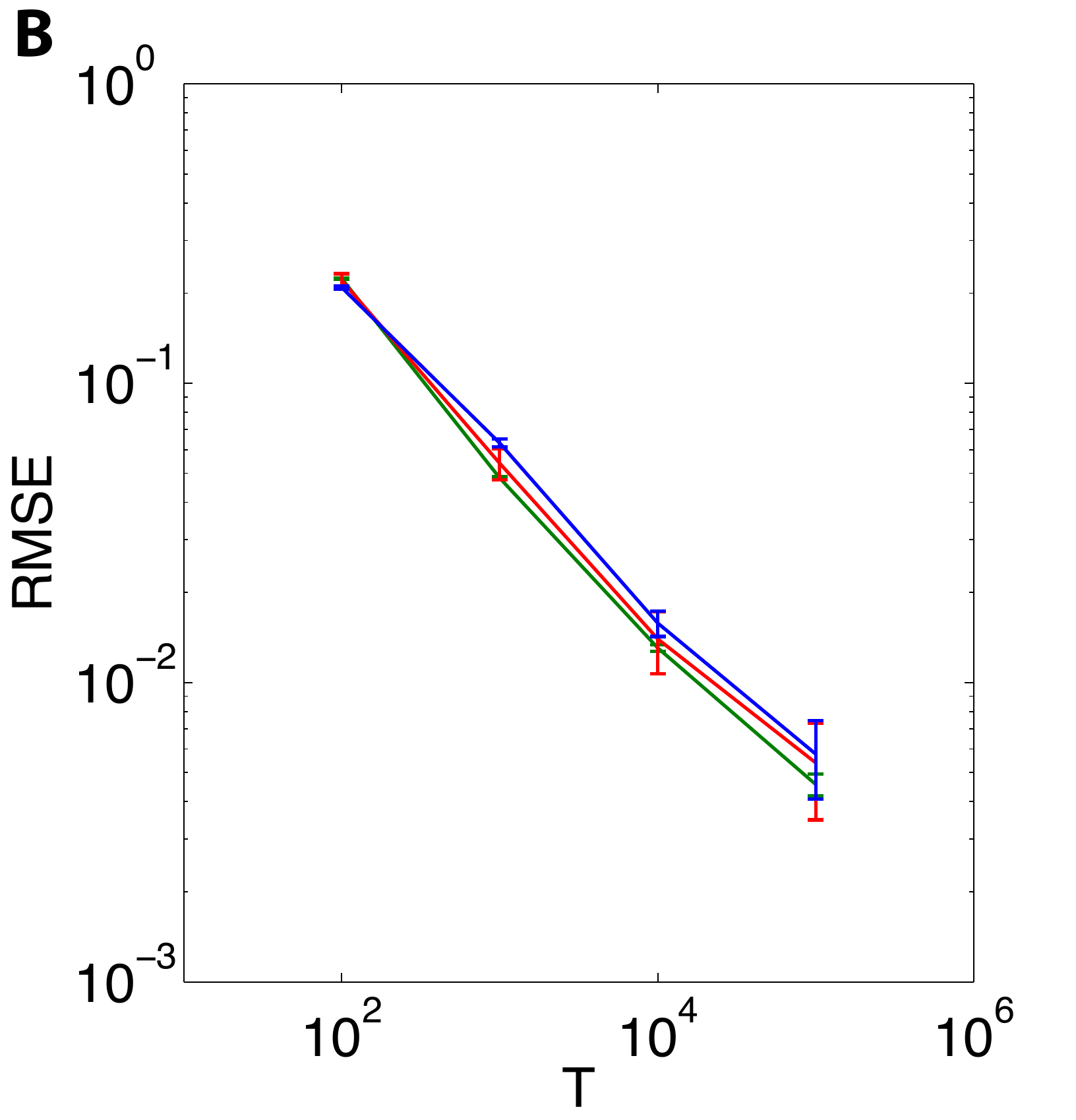}
                \label{fig:RMSEvsTtot(Nv=80,Nh=8)}
                \includegraphics[width=60mm,height=60mm]{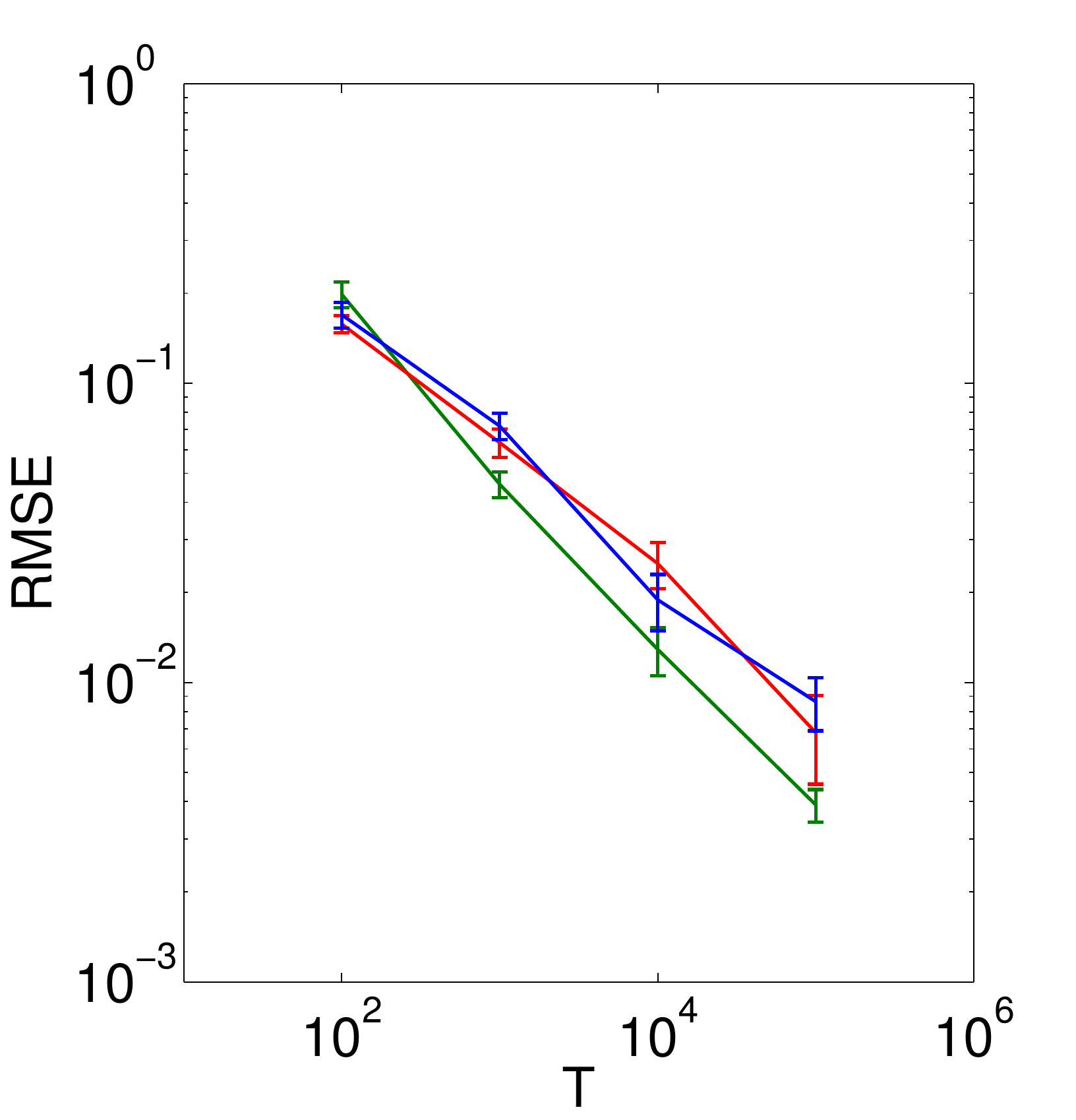}
                \label{fig:RMSEvsT(Nv=80,Nh=16)}
        \caption{Learning on a graphs with $N_{\rm v}=80$ visible units and $N_{\rm h}=8$ hidden units, sparsity c=0.1, g=1. Js (green), Ks (red), Ls(blue). (A) An example showing the reconstructed couplings versus the true ones used in generating the data at a  data length $T=10^5$. Crosses correspond to the initial values for the reconstruction. (B) RMSE on couplings vs data length $T$, $N_{\rm v}=80$, $N_{\rm h}=8$ (left) and $N_{\rm h}=16$ (right). Error bars correspond to one standard deviation on 20 realizations of the generative model. }\label{fig:Scatter}
\end{figure}

We studied the performances of our algorithm in learning the connectivity for sparse and fully connected networks with Gaussian distributed non-zero bounds with standard deviation defined in (\ref{eq:stdJs})-(\ref{eq:stdKs}). Since the system is invariant with respect to permutations of the hidden nodes indices, we initialised our algorithm with initial couplings having the same sign as the true ones. Figure \ref{fig:Scatter}A shows examples of the qualitative agreement between reconstructed $J^{\rm rec}_{ij}$ and generative model couplings $J_{ij}$, while Figure \ref{fig:Scatter}B shows that the RMSE on the reconstructed couplings,

\begin{equation}
{\rm RMSE}\equiv \frac{\sqrt{\sum_{i,j}\left( J^{\rm rec}_{ij}-J_{ij}\right)^2 }}{N_{\rm v}\sigma_{\rm J}}
\end{equation}
scales as $1/\sqrt{T}$ with the data length $(T)$.

\begin{figure}[h]
        \centering
                \includegraphics[width=65mm,height=50mm]{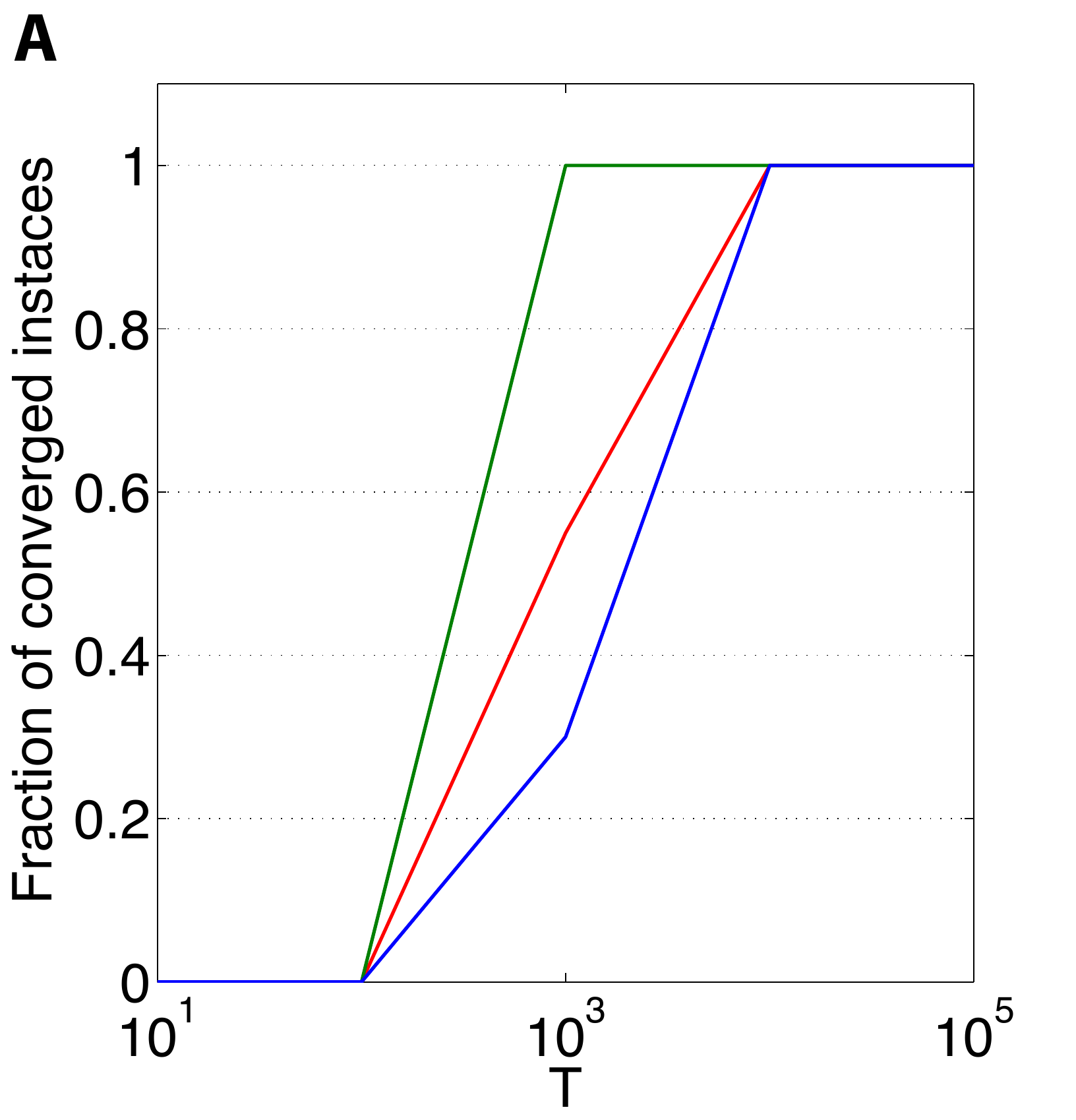}
                \label{fig:CONVvsTbars(c=0.1,g=1.)}
                \includegraphics[width=65mm,height=50mm]{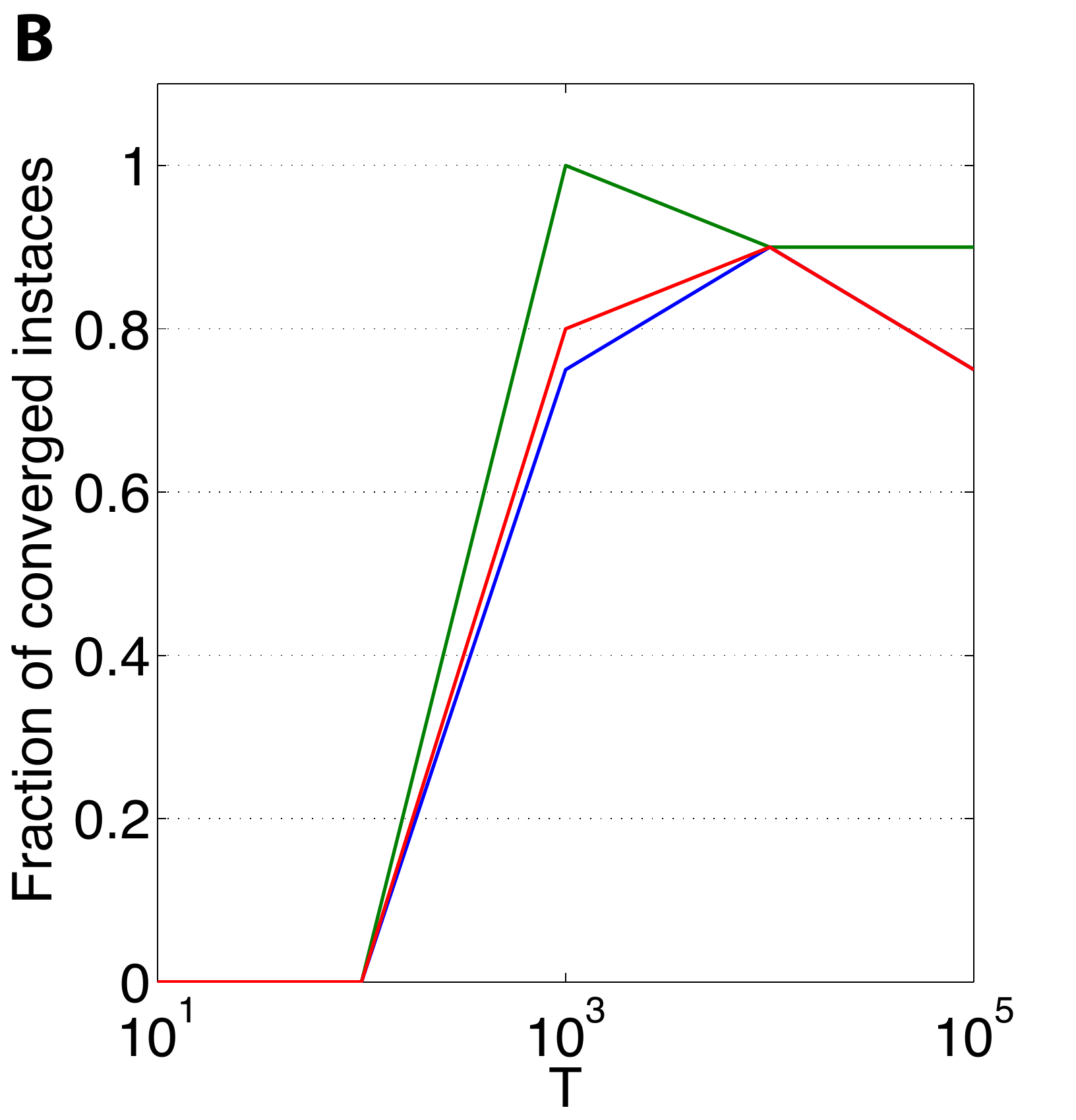}
                \label{fig:CONVvsTbars(c=1.,g=1.)}
                \includegraphics[width=65mm,height=50mm]{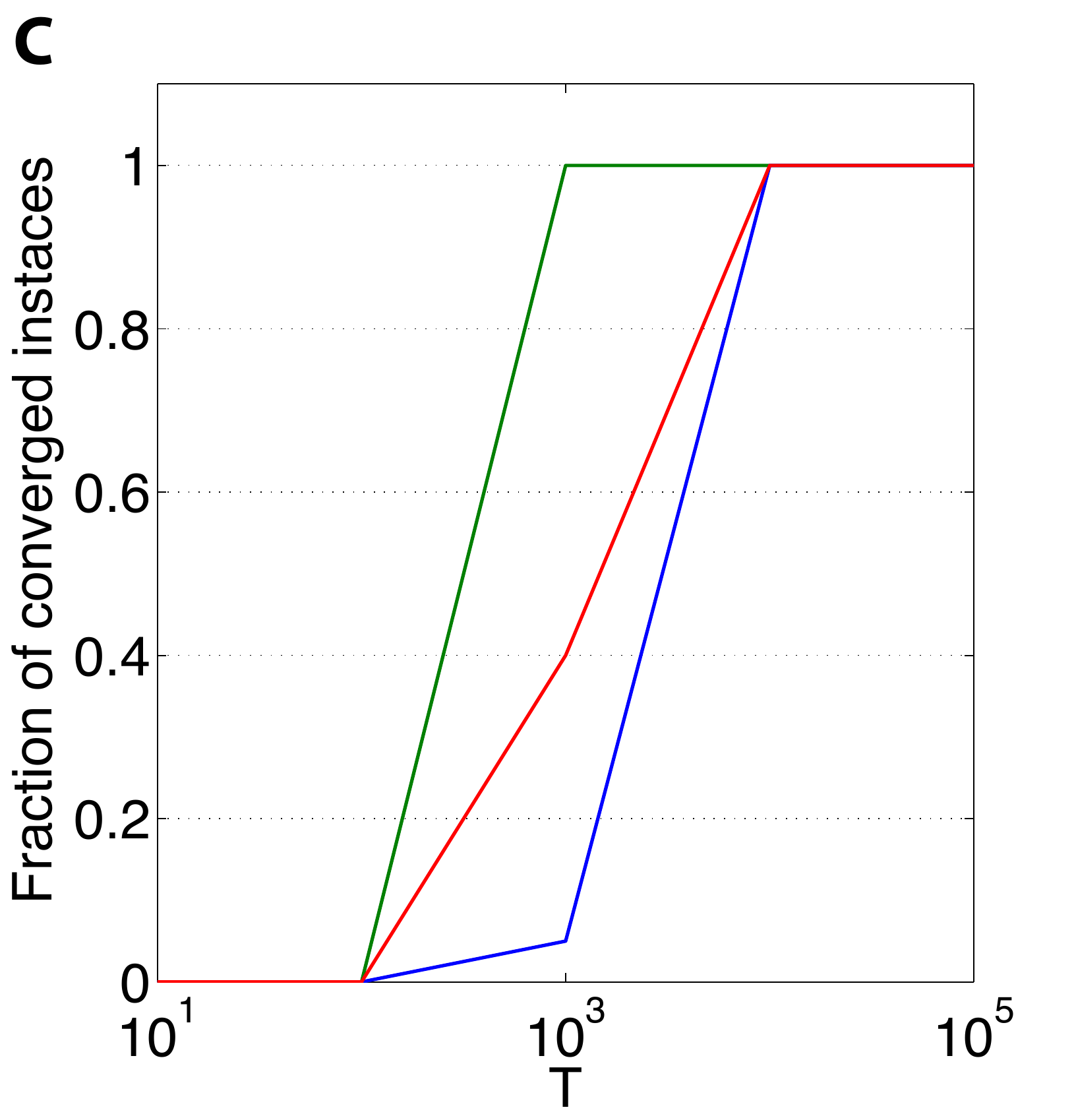}
                \label{fig:CONVvsTbars(c=1,g=0.5)}
                \includegraphics[width=65mm,height=50mm]{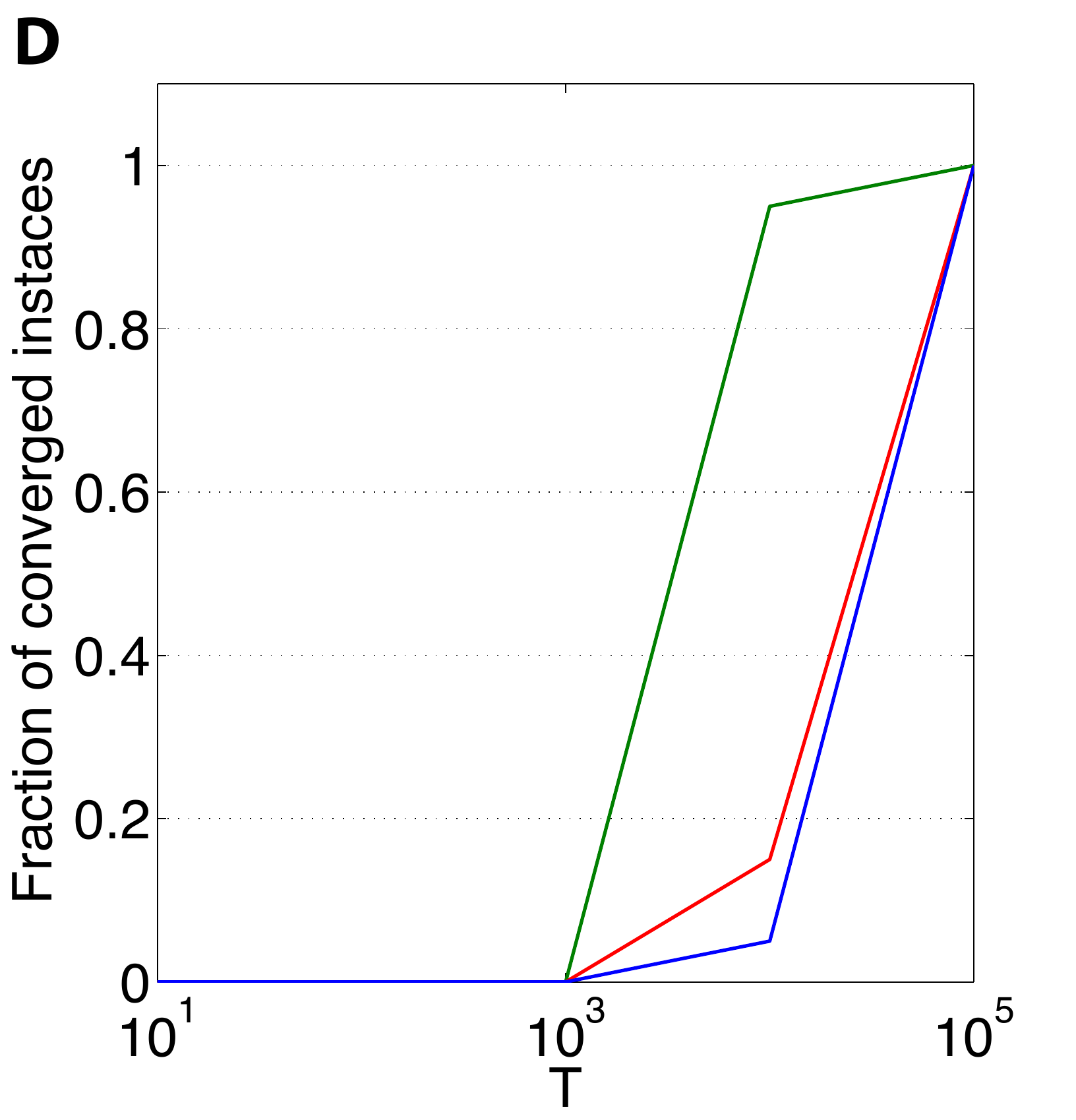}
                \label{fig:CONVvsTbars(c=1,g=0.1)}
        \caption{Fraction of converged instances vs data length for a network of $N_v=20$ visible units and $N_h=4$ hidden. (A) $c=0.1$, $g=1$, (B) $c=1$, $g=1$, (C) $c=1$, $g=0.5$, (D) $c=1$, $g=0.1$. Js (green), Ks (red), Ls(blue) convergence is considered separately.}\label{fig:CONVvsT}
\end{figure}

Figure \ref{fig:CONVvsT} shows that, as expected, the weaker the couplings the longer the data length required for convergence. Figure \ref{fig:CONVvsT} also points out that for fully connected topologies 100\% convergence rate is never reached when the couplings are strong. 

\begin{figure}[h]
        \centering 
             \includegraphics[width=65mm,height=60mm]{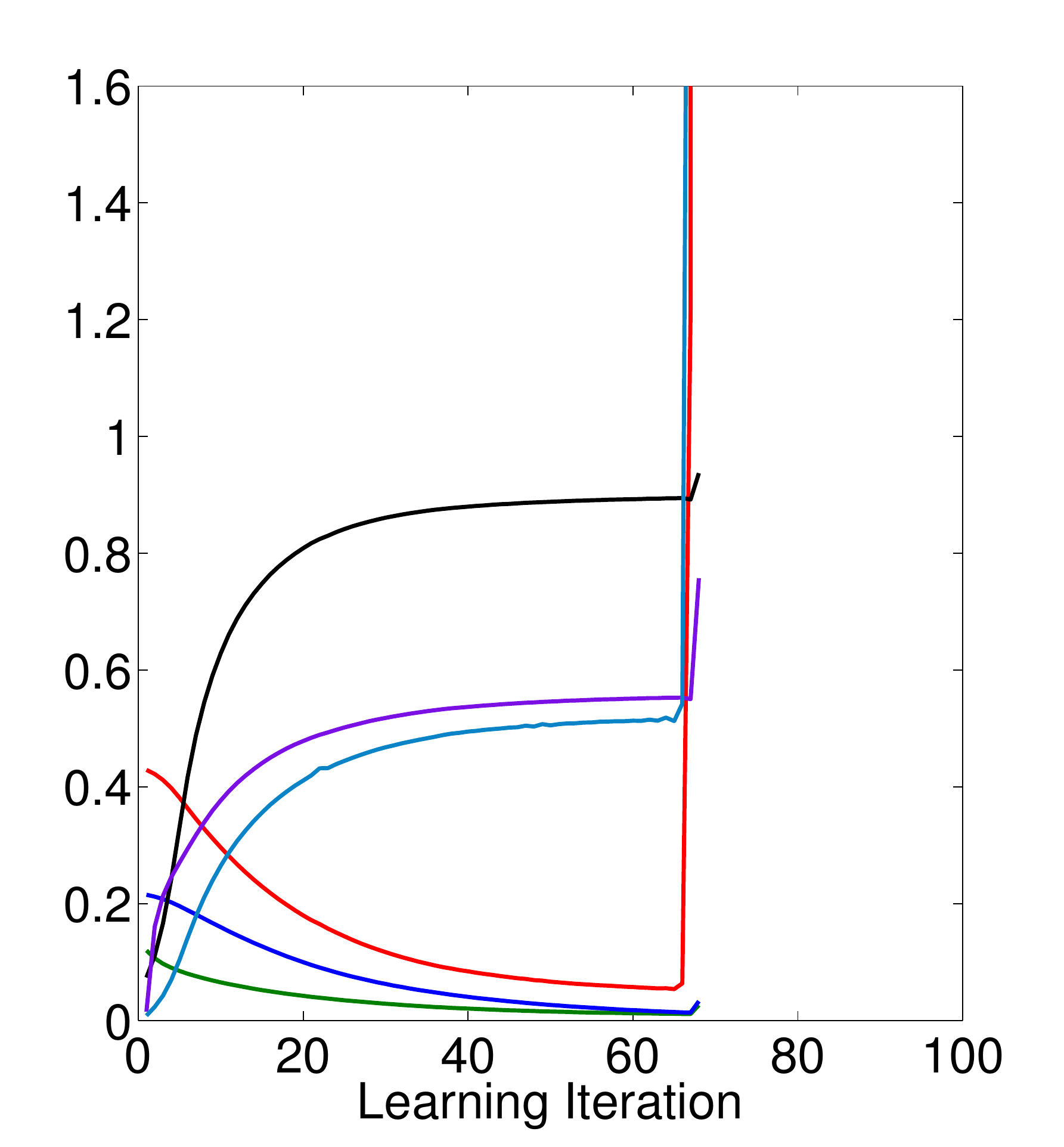}
                \includegraphics[width=65mm,height=60mm]{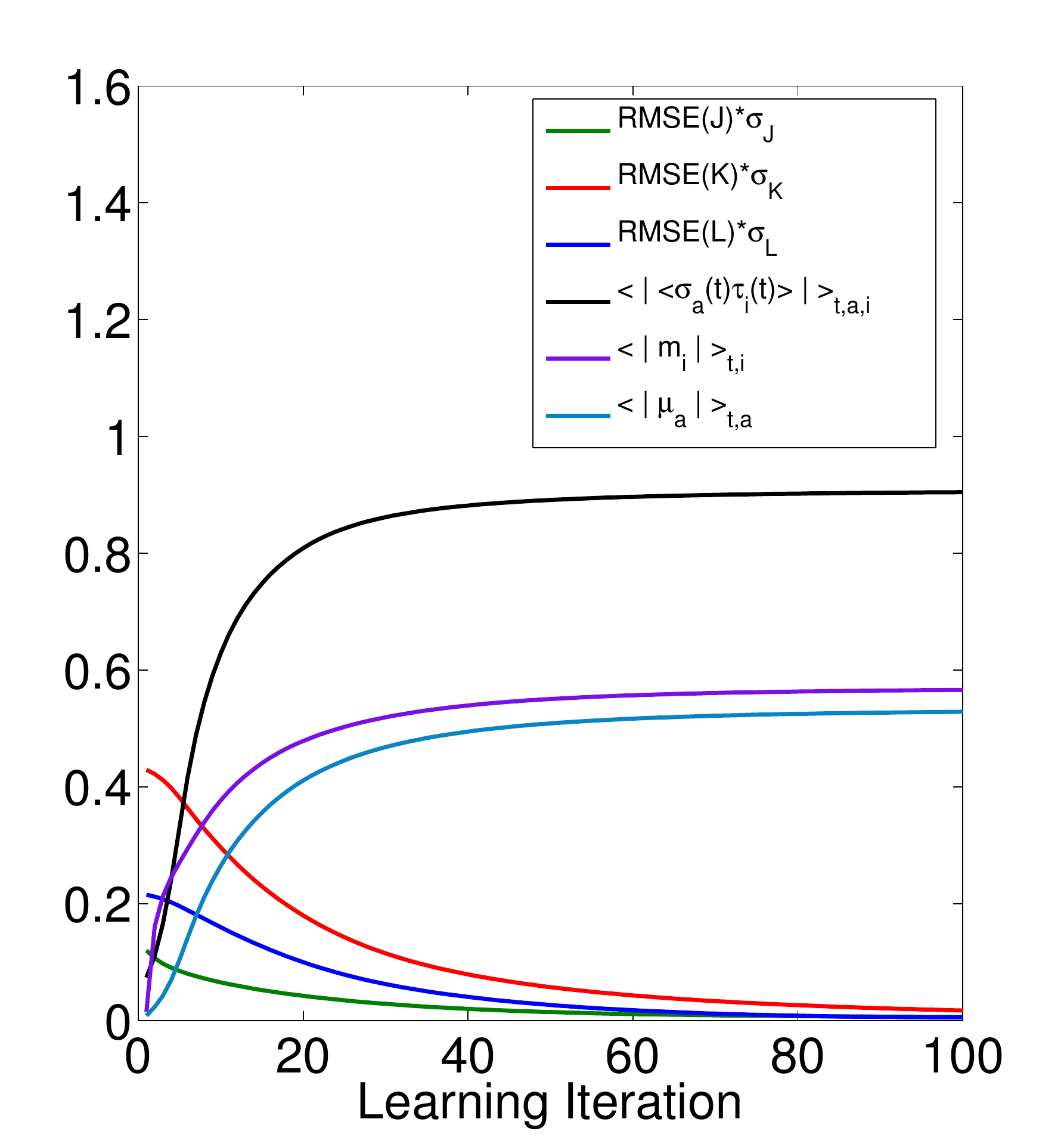}
        \caption{Inference \-- learning vs iteration of our EM algorithm. $T=10^5$, $N_{\rm v}=20$, $N_{\rm h}=4$, $g=1$, $c=1$. RMSE on the couplings and average absolute value of the statistics, during learning of a single connectivity. Left and right plots share the same legend (displayed on the right figure). Left: Uncorrected SusP without correction. Right: corrected SusP. In this example BP equations (\ref{eq:cavityREP1})-(\ref{eq:cavityREP2}) stop converging at the 11th iteration of the algorithm. After few iterations the equations for the correlations (\ref{eq:cavityGRAD1})-(\ref{eq:cavityGRAD2}) follow and our correction starts affecting the learning. Confinement of the inferred correlations ensures the convergence of the learning algorithm.}\label{fig:ShowDiv1TOT}
\end{figure}

\begin{figure}[h]
        \centering
                \includegraphics[width=65mm,height=60mm]{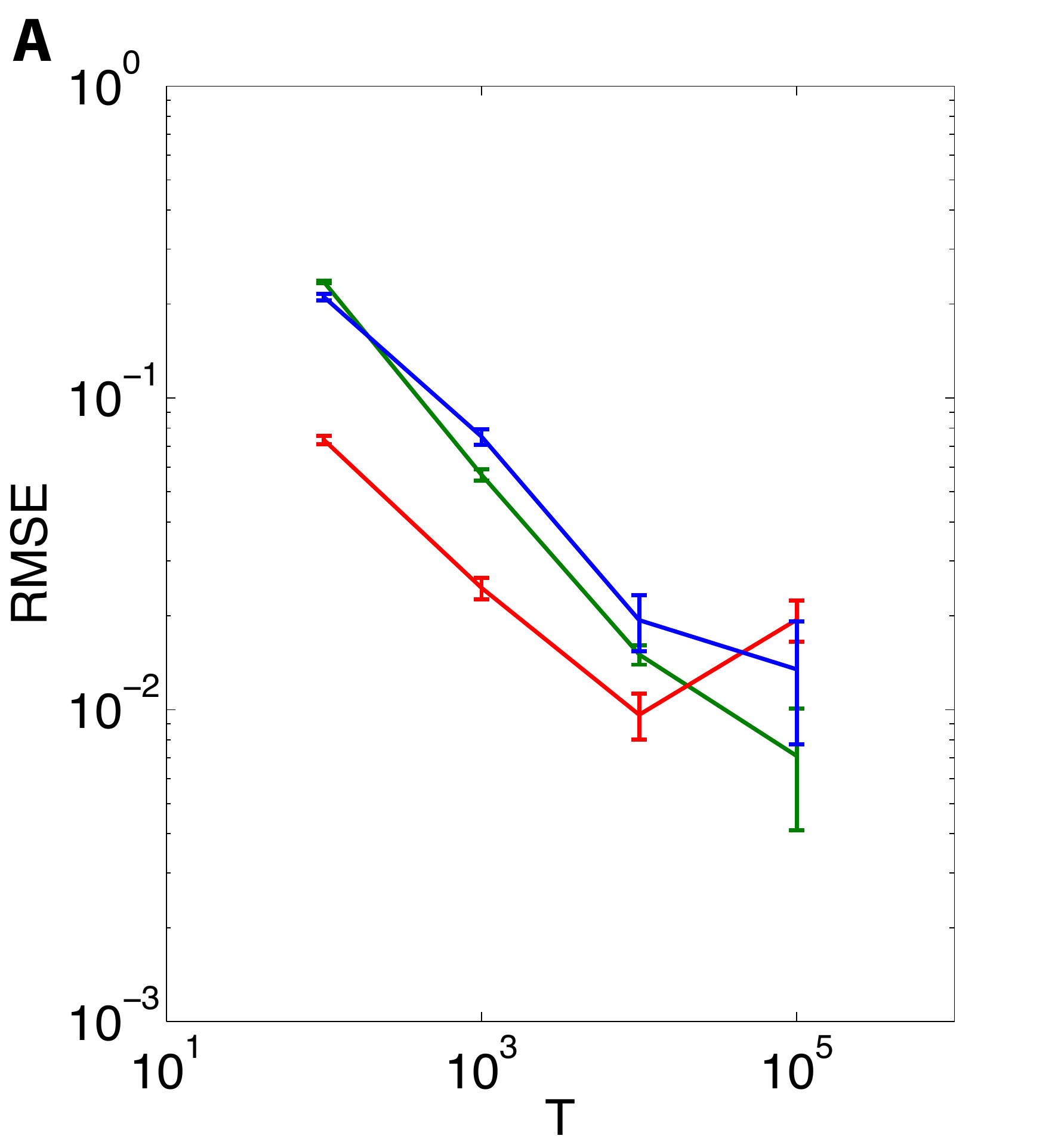}
                \label{fig:NORM(c=0.1)}
                 \includegraphics[width=65mm,height=60mm]{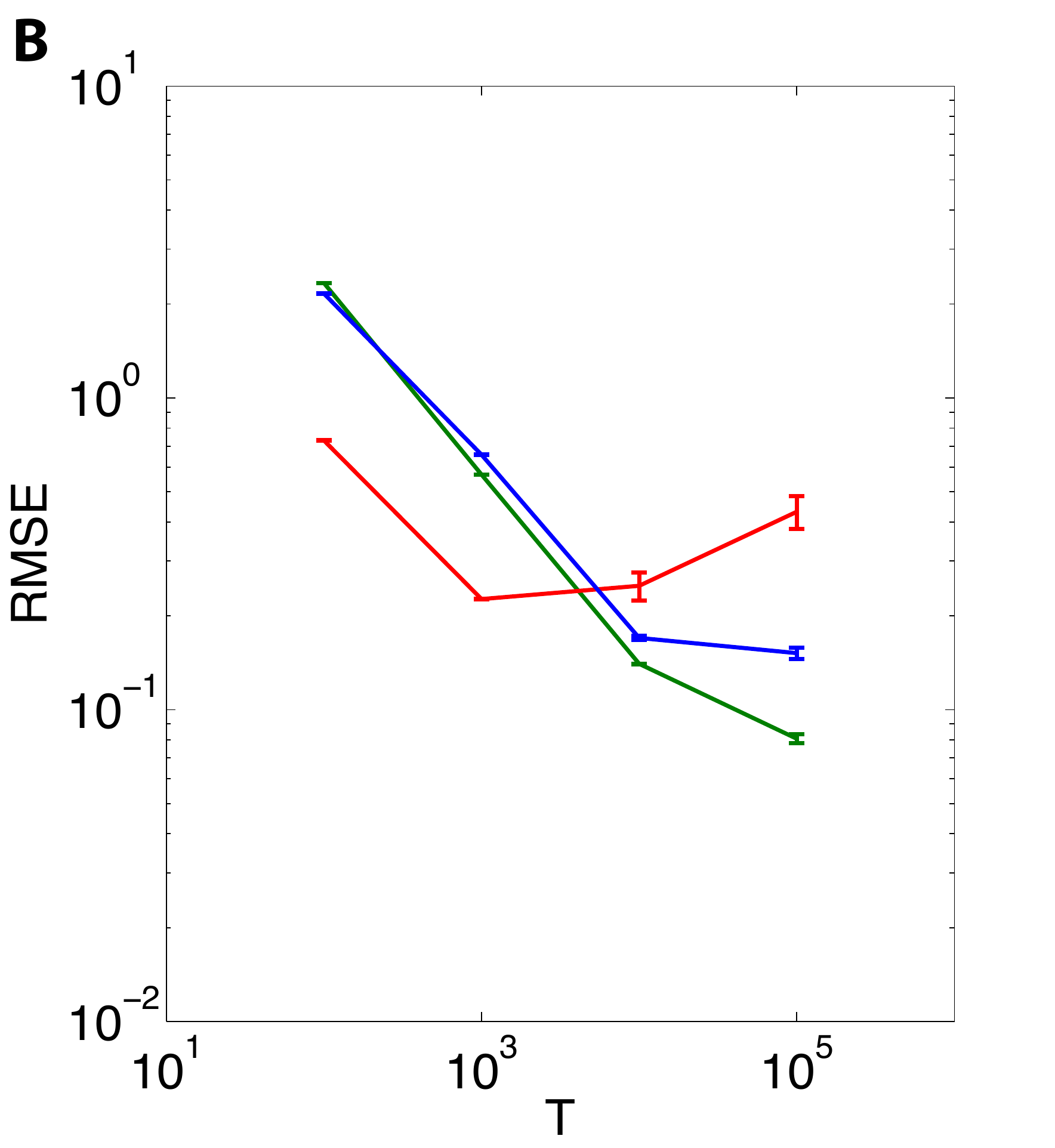}
                \label{fig:NORM(c=1.)}
                 \includegraphics[width=65mm,height=60mm]{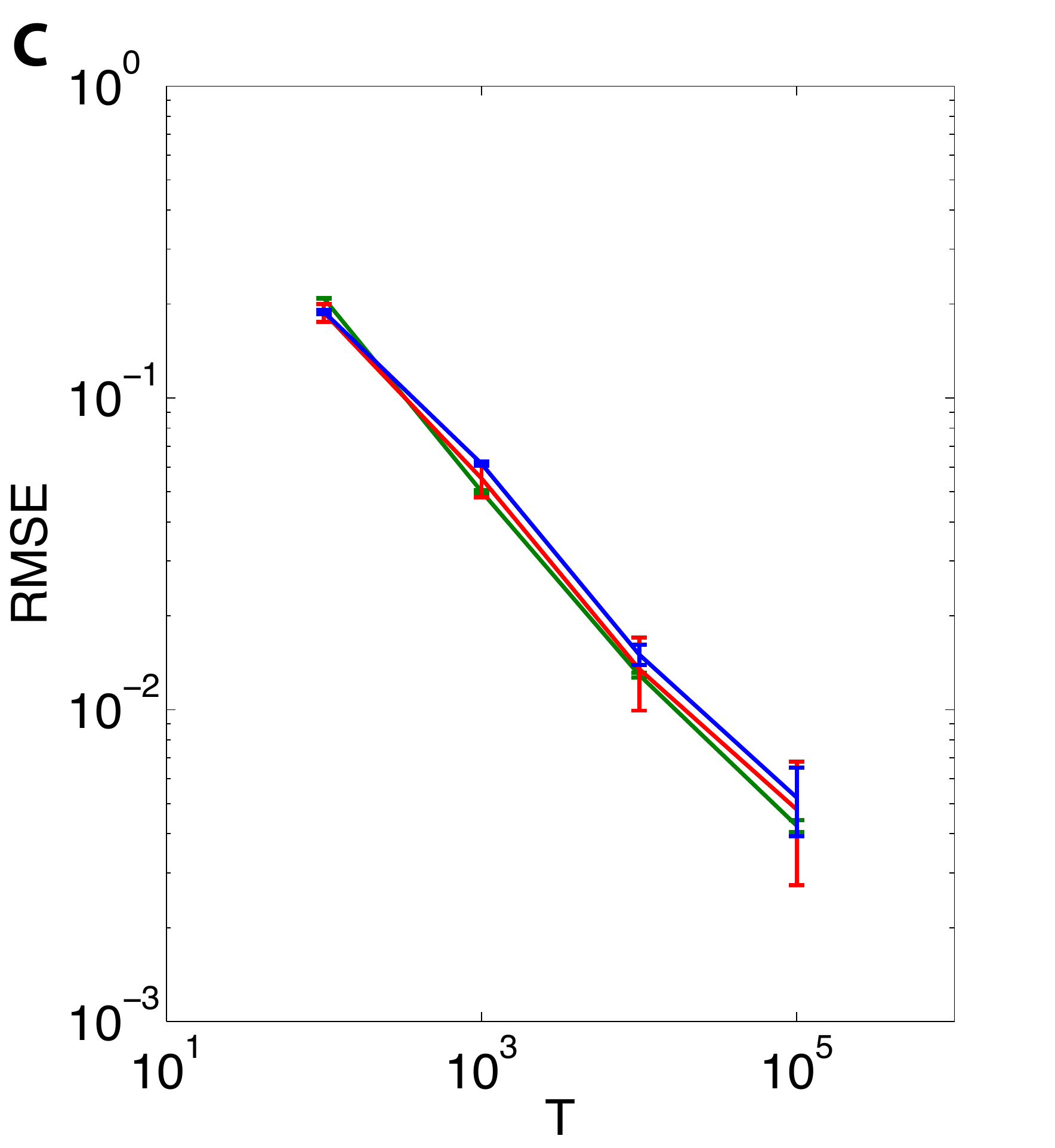}
                \label{fig:BOUNDED(c=0.1)}
                \includegraphics[width=65mm,height=60mm]{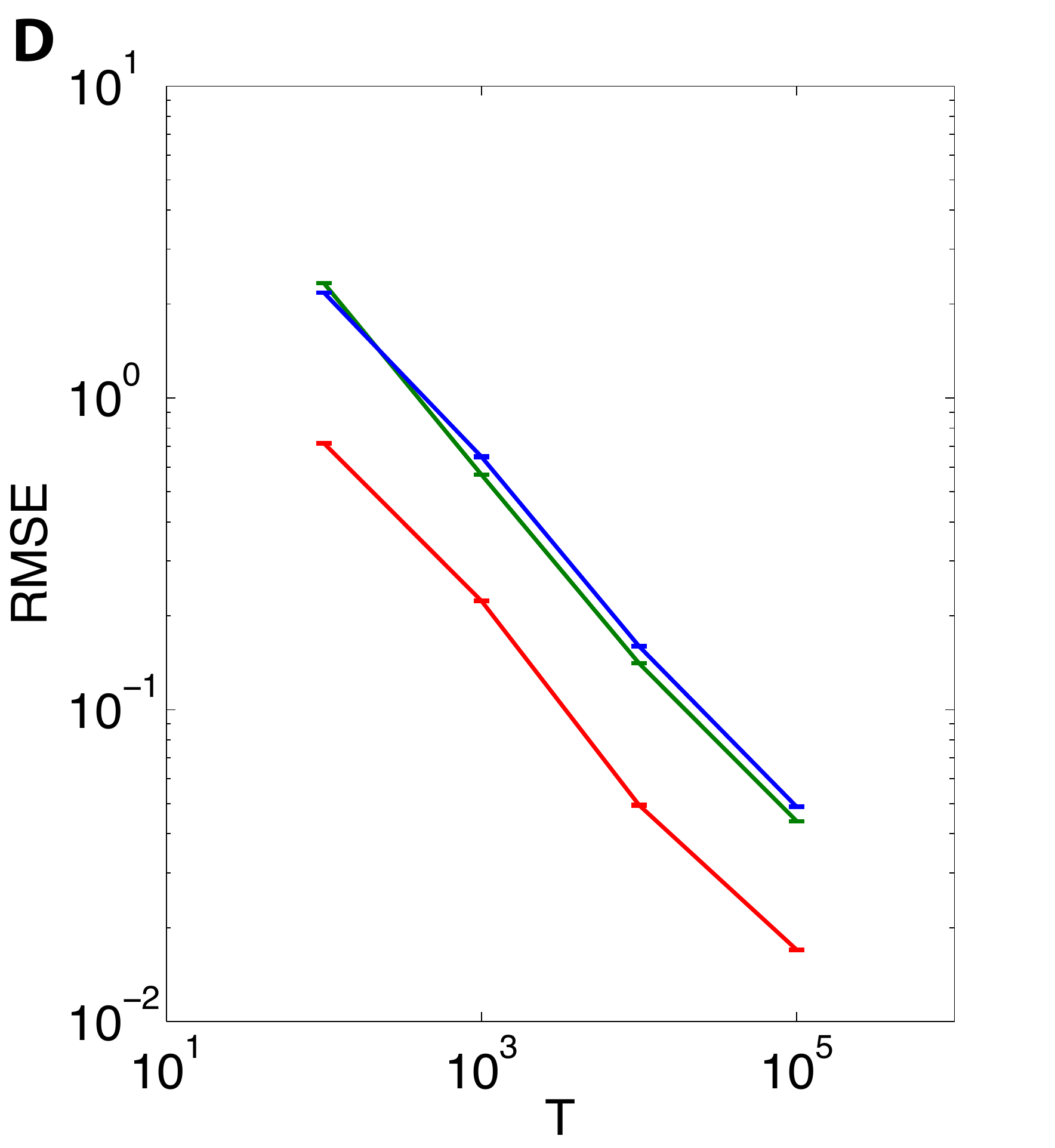}
                \label{fig:BOUNDED(c=1.)}
        \caption{RMSE on the couplings vs data length for networks of $N_{\rm v}=90$ visible units and $N_{\rm h}=10$ hidden, $g=1$. Top: the original algorithm; bottom: the corrected algorithm. (A)-(C) $c=0.1$ (B)-(D) $c=1$. Convergence of Js (green), Ks (red), and Ls(blue) is considered separately. Errors correspond to one standard deviation around the mean for 10 realizations of the couplings. }\label{fig:NORMvsBOUNDED}
\end{figure}

Inspecting the behavior of the algorithm during learning, one realizes that divergence develops in a systematic way. First, after some iterations of the EM-algorithm,  (\ref{eq:cavityREP1})-(\ref{eq:cavityREP2}) for the magnetizations start not converging.  Then, after a few iterations the equations for the correlations (\ref{eq:cavityGRAD1})-(\ref{eq:cavityGRAD2}) stop converging and eventually correlations explode and couplings follow. Fig.\ \ref{fig:ShowDiv1TOT}A displays a typical example of the evolution of the divergence in terms of hidden units statistics and RMSE on the couplings. Notice that we considered both sets of equations, (\ref{eq:cavityREP1})-(\ref{eq:cavityREP2}) and (\ref{eq:cavityGRAD1})-(\ref{eq:cavityGRAD2}), as having converged when the average square distance between cavity magnetizations and correlations at consecutive iterations is smaller than $10^{-9}$ . We have tried to solve the problem starting from its source and improving the convergence of the BP equations for the magnetizations by setting initial values to the inferred value at the previous EM iteration, using  na\"{i}ve mean field theory and initializing $m_i(t)$ at $s_i(t+1)$, as well as damping adaptively the learning. However, we did not observe any significant effect by doing this. We therefore intervened on correlations directly, exploiting the Cauchy--Schwarz inequality to detect diverging two point correlation functions ($\langle \sigma_b \tau_i\rangle>1$) and replacing them by their corresponding independent spin values ($\mu_b m_i$). The correction prevents the correlations from diverging and consequently allows further improvement of the learning as shown in Fig.\ \ref{fig:ShowDiv1TOT}B. Figure \ref{fig:NORMvsBOUNDED} compares RMSE vs data length for the original uncorrected algorithm and the corrected one: improvement of convergence rate affects the quality of the reconstruction as well. 

\begin{figure}[h]
        \centering
                \includegraphics[width=65mm,height=60mm]{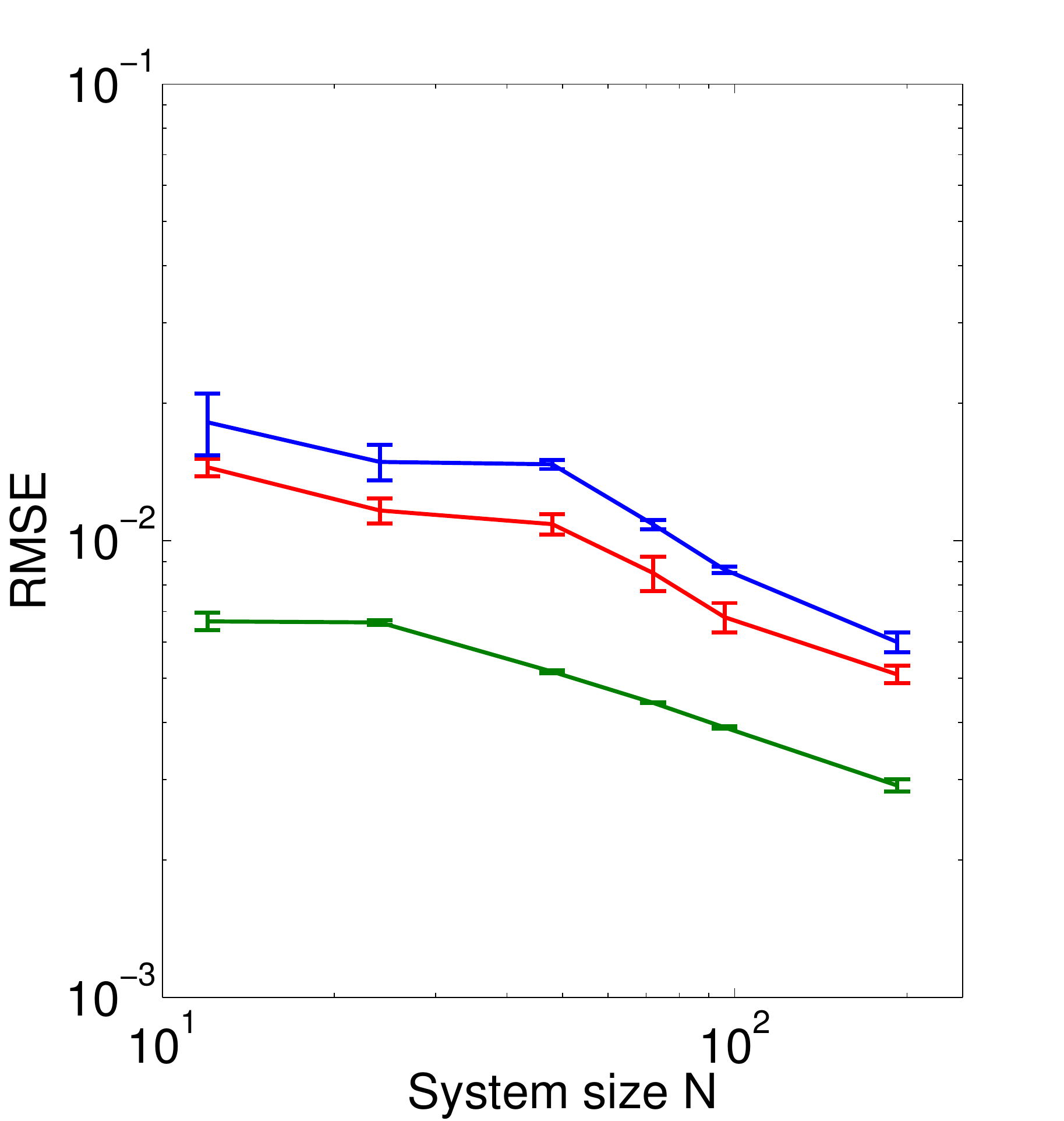}
         \caption{RMSE of the reconstructed couplings vs system size $N$. Data length $T=10^5$, coupling strength $g=1$, sparsity $c=0.1$ and fixed $N_{\rm v}/N_{\rm h}=4$.}
         \label{fig:RMSEvsSIZE}
\end{figure}

Figure \ref{fig:RMSEvsSIZE} shows that the RMSE on the reconstructed couplings decreases with the system size, if the $N_{\rm v}$ to $N_{\rm h}$ ratio and the data length $T$ are fixed. In order to be conclusive on the scaling of the RMSE with the system size, the problem needs further numerical exploration.

\subsection{Binary bonds}

We tested our algorithm on a network with binary-valued connections in addition to the Gaussian-distributed ones that we discussed above. The sparsity is controlled by the parameter $c$, as in the previous section. That is, each bond is zero with probability $c$ and non-zero with probability $1-c$ in which case it can be positive or negative with equal probabilities taking the value

\numparts
\begin{eqnarray}
&J_{ij}=&\pm g/\sqrt{2N_{\rm v} c} \label{eq:binJs}\\
&L_{aj}=&\pm g/\sqrt{N_{\rm v} c} \label{eq:binLs}\\
&K_{ia}=&\pm g/\sqrt{2N_{\rm h} c} \label{eq:binKs}
\end{eqnarray} 
\endnumparts

\begin{figure}[h]
        \centering
             \includegraphics[width=60mm,height=60mm]{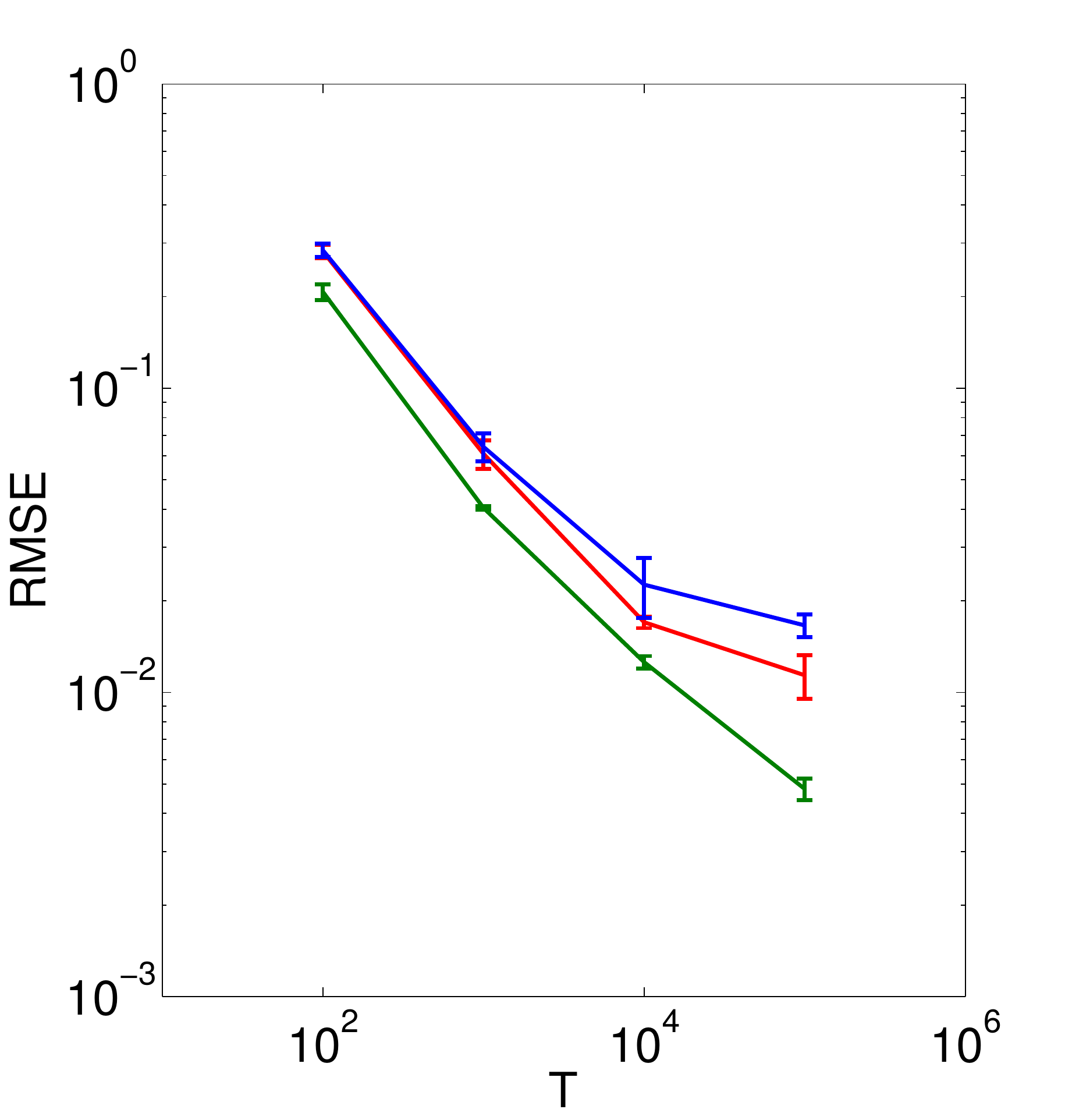}
                \label{fig:Binary(c=01)}
                \includegraphics[width=60mm,height=60mm]{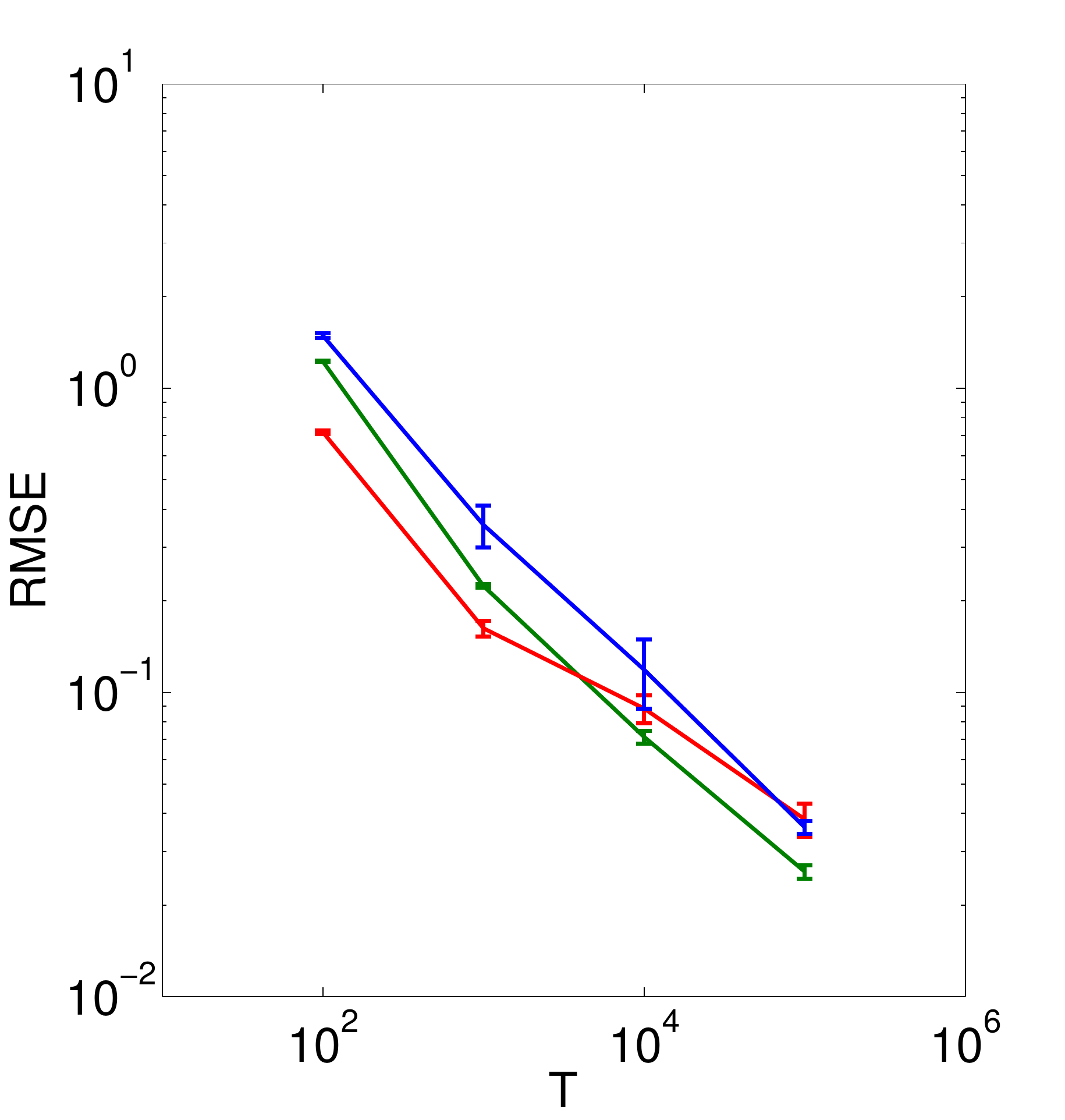}
                \label{fig:Binary(c=1)}
        \caption{Learning on a graphs with binary couplings: RMSE of couplings vs data length. $N_{\rm v}=20$ visible units and $N_{\rm h}=4$ hidden units, sparsity $c=0.1$ (left) and $c=1$ (right). Js (green), Ks (red), Ls(blue).}\label{fig:Binary}
\end{figure}

Figure \ref{fig:Binary} shows how the RMSE scales with the data length for dilute and fully connected systems.

In order to investigate how much the previous knowledge of the degree of sparsity affects the reconstruction, we attempted an {\em a posteriori} pruning of the reconstructed couplings. The latter consisted of a classification of the reconstructed couplings according to their strength as zero and non-zero couplings, consistent with the sparsity. The first group are set to zero, while remaining couplings are sorted by sign, and then set to their respective inferred means. Unfortunately the process did not improve the quality of the reconstruction due to misclassification: the distribution of the reconstructed couplings is not trimodal and the three distributions (zero, positive and negative couplings) overlap.

\subsection{Comparison with TAP}

\begin{figure}[h]
        \centering
             \includegraphics[width=60mm,height=60mm]{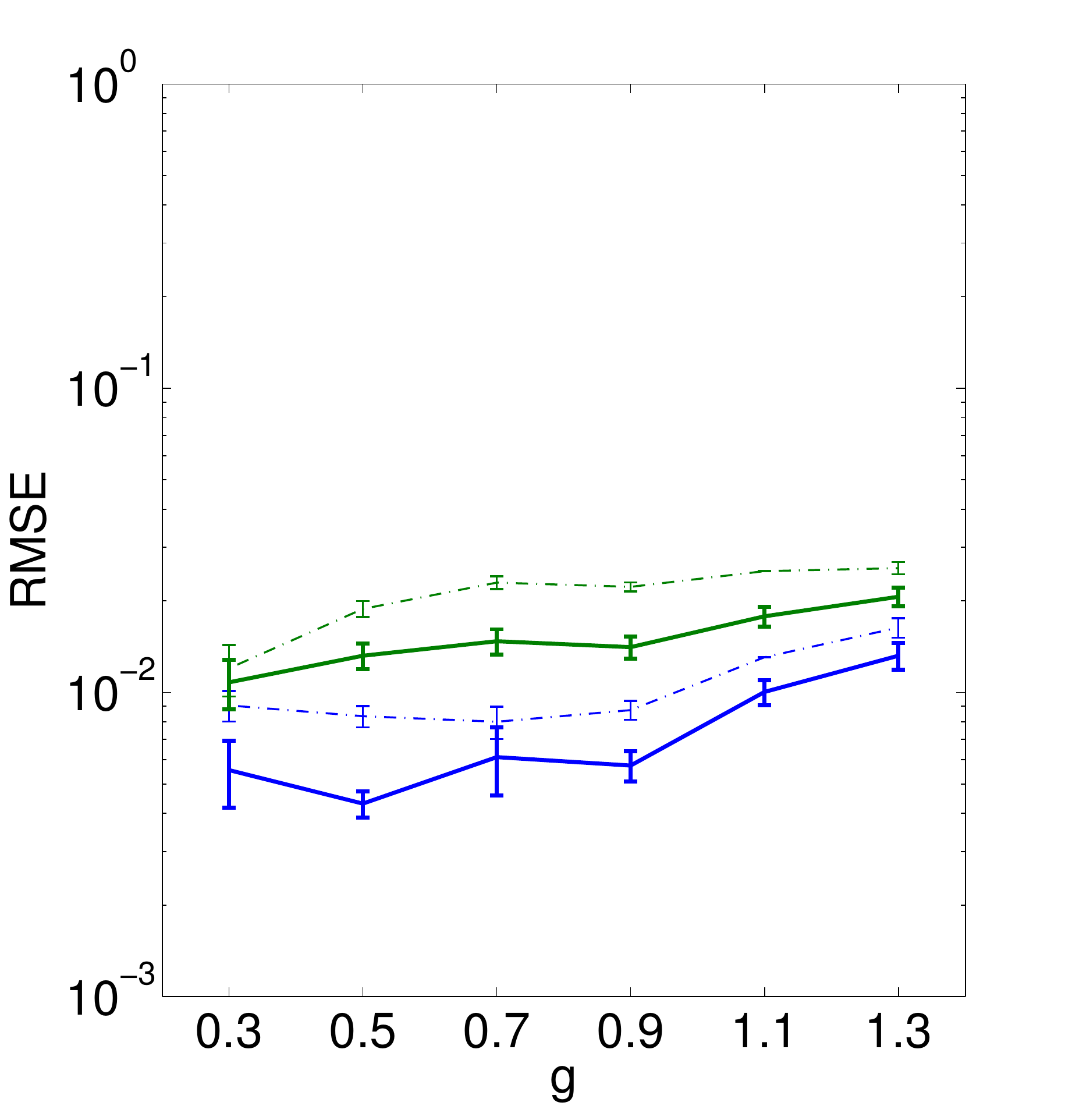}
                \label{fig:TAP-BP(c=01)}
                 \includegraphics[width=60mm,height=60mm]{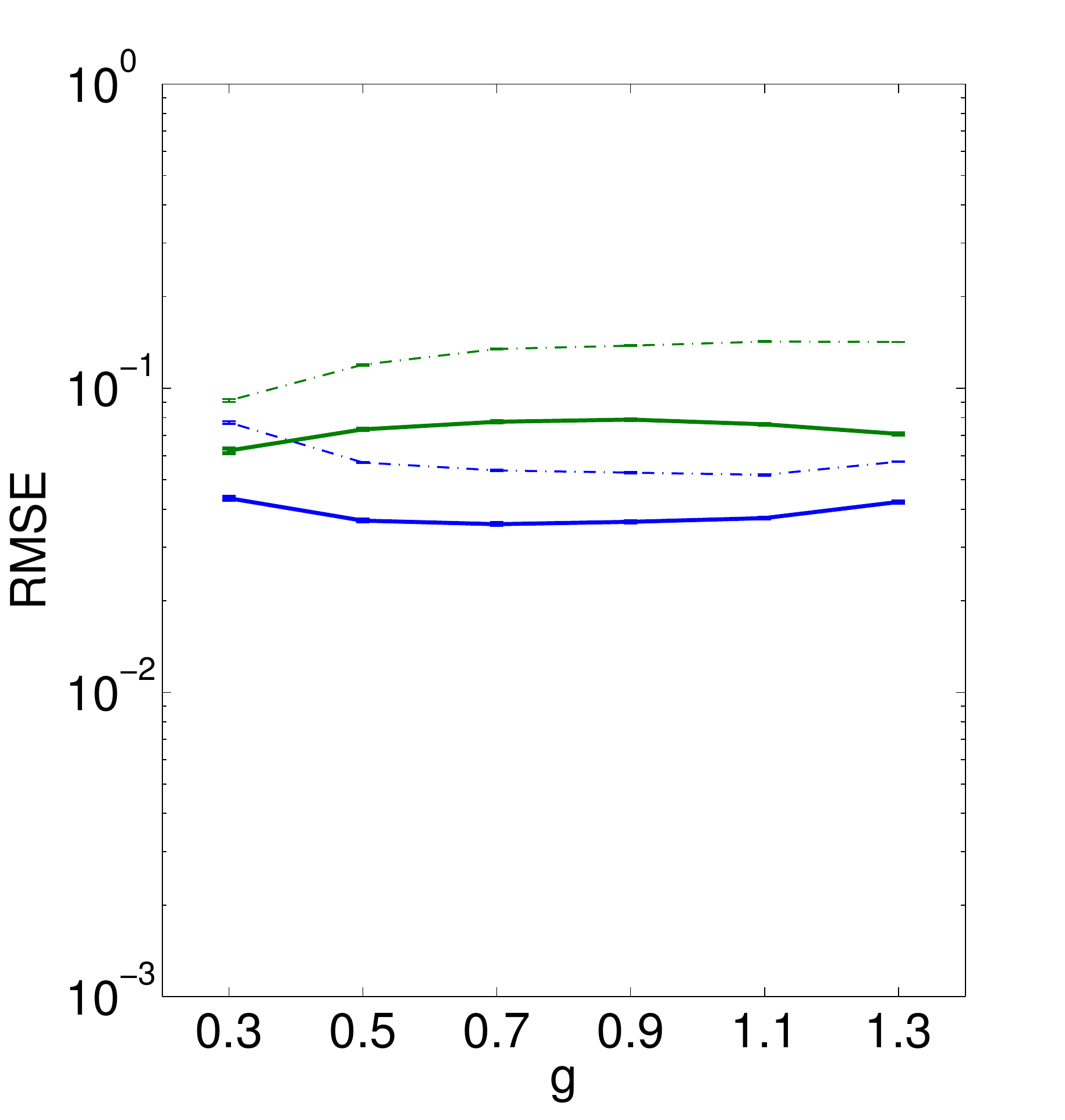}
                \label{fig:TAP-BP(c=1)}
        \caption{BP-TAP comparison: RMSE on the reconstructed couplings vs couplings strength g. Solid (dashed) lines correspond to $N_{\rm v}=90$ ($N_{\rm v}=80$) visible units and $N_{\rm h}=10$ ($N_{\rm h}=20$) hidden units, sparsity $c=0.1$ (left), $c=1$ (right), TAP (green), BP (blue), $T=10^5$. Error bars correspond to one standard deviation on 10 realizations of the generative model. }\label{fig:TAP-BP}
\end{figure}

We also compared the performance to that obtained for the algorithms recently developed in \cite{dunn2013learning,hertz2014Network} . In the absence of hidden-to-hidden interactions and for weak couplings, the authors of those papers derived the same set of TAP-like equations: in the first paper from the saddle point approximation to the path integral of the likelihood in equation (\ref{eq:factLike}) and in the second through a variational approach.  We find here that BP outperforms TAP both on fully connected and sparse topologies, as can be seen in Figure \ref{fig:TAP-BP}. Here we plot the RMSE of the reconstructed couplings as a function of the coupling strength $g$.  

\newpage

\section{Discussion}{\label{sec:Discussion}}

In this paper we presented a novel approach for studying the dynamics of a kinetic Ising model with hidden nodes and learning the connections in such a network. For a system without hidden-to-hidden connections, the likelihood of the data can be written as a set of independent equilibrium problems for each time step. In each such problem, calculating the partition function requires performing a trace over hidden variables. We showed that the form of the action we get allows performing this trace at the cost of introducing auxiliary replicated spins that mediate the interaction between hidden and observed nodes at the same time. Combined further with Belief Propagation and Susceptibility Propagation, we derived learning rules for inferring the state of the hidden nodes and the couplings in the network, given the observed data. 

The algorithm we developed here was derived in the replica symmetric framework, which may make one wonder whether a form of replica symmetry breaking might be introduced in our approach. However, it is important to note that due to the lack of terms of the form $\tau^{\alpha} \tau^{\beta}$, there is no interaction between the replicas and thus no possibility of breaking the replica symmetry. Consequently within the limit where the number of replicas goes to $-1$, the replica symmetric solution is the true solution. 

We assessed the accuracy of our approximate method in both estimating the hidden units' statistics and solving the inverse problem. The choice of the BP algorithm for the inference step was motivated by its good performance on equilibrium spin glasses \cite{ricci2012bethe}.  We noticed that when trying to learn the connections, for which we used the SusP algorithm, the instability of that algorithm affects the reconstruction of strong couplings. We thus suggested a simple but effective correction to improve convergence, finding that it ensured that the RMSE on the couplings scales as $1/\sqrt{T}$. Finally, we compared our algorithm to the TAP approach \cite{dunn2013learning}, finding that the algorithm described here systematically outperforms the latter, though by a small margin. Although the algorithm proposed here improves on TAP even for weak couplings, it is important to note that TAP is polynomial in the number of hidden nodes, while our algorithm is polynomial in the total system size. Furthermore, TAP learning allows for the reconstruction of the couplings in the presence of hidden-to-hidden connections. It would therefore be interesting to combine the two algorithms in the following way: during learning one can initially set the hidden-to-hidden couplings to zero, learning all the other connections using the BP-replica method proposed here, then moving to learn the hidden-to-hidden couplings using the TAP equations. 

For the case of sparse networks, there are some modifications of the algorithm that we have not yet tried, but they may improve the method. For instance it would be interesting to try an online decimation, that is forcing the known sparsity at every step of learning, instead of the off-line one that we tried. It is also well known that, particularly for small data sets, one can mistakenly infer that absent connections are present but with small values. Thus it would be interesting to test non-trivial standard methods for optimally pruning the reconstructed networks, like $L_1$ regularization. 

\section*{Acknowledgments}
This work has been partially supported by the Marie Curie Initial Training Network NETADIS (FP7, grant 290038)
\section*{References}

\bibliography{Bibliography}{}
\bibliographystyle{iopart-num}

\end{document}